\documentclass{article}

\usepackage{PRIMEarxiv}

\usepackage[utf8]{inputenc} 
\usepackage[T1]{fontenc}    
\usepackage{hyperref}       
\usepackage{url}            
\usepackage{booktabs}       
\usepackage{amsfonts}       
\usepackage{nicefrac}       
\usepackage{microtype}      
\usepackage{lipsum}
\usepackage{fancyhdr}       
\usepackage{graphicx}       
\usepackage{listings}
\usepackage{algorithm}
\usepackage{algpseudocode}
\usepackage{hyperref}
\usepackage{currfile}
\usepackage{amsmath}
\usepackage{amsthm}
\usepackage[algo2e]{algorithm2e} 

\usepackage{subcaption}
\usepackage{wrapfig}
\usepackage{amssymb}

\newtheorem{theorem}{Theorem}[section]
\newtheorem{example}{Example}[section]
\newtheorem{definition}{Definition}
\newtheorem{proposition}{Proposition}
\newtheorem{remark}{Remark}

\graphicspath{{media/}}     

\title{Polynomial Invariant Generation for Floating-Point Programs
}

\author{
  CAI, Xuran \\
  University of Oxford \\
  xuran.cai@cs.ox.ac.uk\\
   \And
  CHEN, Liqian \\
  National University of Defense Technology\\
  lqchen@nudt.edu.cn\\
 \And
 FU, Hongfei\\
 Shanghai Jiao Tong University\\
 jt002845@sjtu.edu.cn\\
}

\begin{document}
\maketitle

\begin{abstract}
In numeric-intensive computations, it is well known that the execution of floating point programs is imprecise as floating point arithmetic incurs round-off errors. Although round-off errors are small for a single floating point operation, the aggregation of such errors may be dramatic and cause catastrophic program failures. Therefore, to ensure the correctness of floating point programs, round-off error needs to be carefully taken into account. In this work, we consider polynomial invariant generation for floating point programs, aiming at generating tight invariants under the perturbation of round-off errors. Our  contribution is a novel framework for applying polynomial constraint solving to address the invariant generation problem, which is also the first polynomial constraint solving based approach that handles floating point errors to our best knowledge. In our framework, we propose a novel combination of round-off error analysis and polynomial constraint solving, aiming to circumvent the  cost of handling a large number of error variables in the floating point model. Experimental results over a variety of challenging benchmarks show that our framework outperforms SOTA approaches in both time efficiency and the precision of generated invariants. 
\end{abstract}

\section{Introduction}

In floating-point programs, floating-point arithmetic introduces round-off errors due to the finite bit representation of floating-point numbers in hardware. Although individual round-off errors are small, their accumulation can lead to significant deviations and unexpected program failures. To ensure correctness despite these errors, formal analysis methods that account for round-off errors have been extensively studied. Various approaches based on different techniques (including abstract interpretation~\cite{BCC+03,Min04,Gou01,GMP02}, optimization methods~\cite{fptaylor,DasBGKP20,MagronCD17}, and data-driven analysis~\cite{pine}) have been proposed.

The primary focus in formal analysis of floating-point programs is \emph{round-off error analysis}, which aims to establish guaranteed bounds on the deviation between the outputs of programs with and without round-off errors. Abstraction-based approaches, such as FLUCTUAT \cite{Gou01,GMP02} and PRECiSA \cite{MoscatoTDM17,TitoloFMM18,TitoloMFMM24}, compute sound over-approximations of accumulated round-off errors. Optimization-based methods, like FPTaylor~\cite{fptaylor} and Satire \cite{DasBGKP20}, derive mathematical characterizations of round-off errors to solve for deviation bounds.

Another important question is invariant generation for floating-point programs, which seeks to produce tight invariants that account for round-off errors. This question is orthogonal to round-off error analysis, as it focuses on generating assertions that over-approximate program behavior under round-off errors, while round-off error analysis derives upper bounds on deviations.

Despite its practical importance, invariant generation 
for floating-point programs 
has received comparatively less attention. Early work such as \emph{ASTR\'EE}~\cite{BCC+03} applies abstract interpretation to infer invariants by soundly abstracting floating-point expressions. Recent approaches include TVPI-FP~\cite{tvpifp}, which uses a two-variable affine abstract domain, and PINE~\cite{pine}, which adopts a data-driven approach that samples program executions and encloses reachable program states with ellipsoidal assertions. Moreover, most existing invariant generation methods treat floating-point computations as exact real arithmetic (i.e.,  ignoring round-off errors), so that they are not sound when directly applied to floating-point programs. Therefore, to generate invariants that remain valid under round-off errors imposes new challenges against the existing literature.

\smallskip
\noindent\emph{Our contributions.} In this work, 
we propose a novel framework to address the problem of generating polynomial invariants for floating-point programs. Our framework has a novel integration between existing polynomial constraint solving methods~\cite{DBLP:conf/pldi/AsadiC0GM21,DBLP:conf/cav/ColonSS03,DBLP:conf/sas/SankaranarayananSM04,polyqent,DBLP:journals/pacmpl/LiuFYSL22,DBLP:journals/toplas/WuWXZZY25} (that do not consider round-off errors) and round-off error analysis methods~\cite{fptaylor,MoscatoTDM17,DasBGKP20} (that do not focus on invariants). 

The integration is nontrivial as we show that special care needs to be taken to ensure the soundness. The detailed contributions are listed as follows.

First, we propose a framework to tackle the practical case
where floating-point variables in the program take bounded values.
Note that most floating-point programs execute with variables falling in a bounded range, while programs involving unbounded values cause overflow exceptions and lead to unbounded round-off errors that are hard to analyze.

The framework iteratively guesses bounded ranges to enclose reachable program states and uses the guessed ranges as mandatory input range to apply round-off error analysis. The derived round-off bounds are then used in the polynomial constraint solving. The framework outputs a polynomial invariant only when the guessed ranges are implied by the invariant, which is crucial to guarantee the soundness. 
The key insight to integrate round-off error analysis is that it eliminates a potentially large number of error variables from the floating-point model to ease the polynomial constraint solving.    

Then, we instantiate our framework with existing polynomial solving methods~\cite{DBLP:conf/pldi/AsadiC0GM21,DBLP:conf/cav/ColonSS03,DBLP:conf/sas/SankaranarayananSM04,polyqent,DBLP:journals/pacmpl/LiuFYSL22,DBLP:journals/toplas/WuWXZZY25}, while enhancing them with techniques such as 
barrier certificates, optimization methods and numerical reparation.

After that, we evaluated our framework on a variety of benchmarks involving complex polynomial and division computations from the literature, demonstrating its effectiveness in generating invariants for a diverse range of floating-point programs when compared with SOTA approaches.

Finally, we discuss an alternative to our framework that circumvents the iterative guess of bounded ranges, at the cost of the limitation to polynomial floating-point programs without division.
\section{Preliminaries}\label{sec:pre}


We define valuations to be used throughout the paper. A \emph{valuation} over a finite set $V$ of variables is a function $\mathbf{v}: V\rightarrow \mathbb{R}$ that assigns a real value $\mathbf{v}(x)$ to every variable $x$ of $V$.
The satisfaction relation $\models$ between valuations and logical formulas over numerical variables is defined in the standard way: a valuation $\mathbf{v}$ over variables $V$ is said to satisfy a formula $\phi$ with variables from $V$, written as $\mathbf{v}\models \phi$, if $\phi$ is true under the substitution (or interpretation) that assigns each variable $x\in V$ the value $\mathbf{v}(x)$.
We distinguish between two types of variables: \emph{program} variables, which are variables appearing in a program, and \emph{error} variables, which represent the round-off errors in a floating-point model.   A \emph{program valuation} is a valuation over a finite set of program variables, while an \emph{error valuation} is a valuation over a finite set of error variables that assigns to every error variable a feasible deviation value. We always assume an implicit linear order among variables so that a valuation can be treated as a vector.  

Below we describe the standard IEEE 754 floating-point model,  the first-order differential characterization for round-off error analysis by FPTaylor~\cite{fptaylor}, and the syntax and semantics of floating-point programs considered in this work.

\subsection{Floating-Point Model}\label{sect:fpmodel}

Modern computers use floating-point representations to represent a finite subset of the real numbers, among which the IEEE 754 floating-point standard \cite{IEEE754} is the most commonly used floating-point representation. 
In the IEEE 754 standard, the binary representation of a floating-point number is  described as $(-1)^S\times M\times 2^E$ where $S\in\{0,1\}$ is the $1$-bit \emph{sign} of the number, 
$M=m_0.m_1m_2\ldots m_\mathbf{p}$ is called the \emph{significand} (wherein $.m_1m_2\ldots m_\mathbf{p}$ represents a  $\mathbf{p}$-bit fraction and $m_0$ is the hidden bit), and $E=e-\mathbf{bias}$ is called the \emph{exponent} (wherein $e$ is a biased $\mathbf{e}$-bit unsigned integer and $\mathbf{bias}=2^{\mathbf{e}-1}-1$). 
The values of $\mathbf{e},\mathbf{bias},\mathbf{p}$ depend on the specific floating-point format. For example, for the 32-bit single-precision format, we have $\mathbf{e}=8$ (and thus $\mathbf{bias}=127$), $\mathbf{p}=23$. 

When a real number cannot be precisely encoded by a floating-point number (with finite bits), the IEEE 754 standard provides different rounding modes (including toward nearest, toward +$\infty$, toward -$\infty$, 
and toward zero) to convert a real number to a nearby floating-point number. Also, the result of a floating-point operation over floating-point numbers may not be exactly representable in the floating-point representation, and thus the result also needs to be rounded into a floating-point number. Let $x_\mathbf{f}$ denote the floating-point representation of a real number $x$ using floating-point format $\mathbf{f}$. The \emph{rounding} error (also called  \emph{round-off} error) is $x-x_\mathbf{f}$. 
There are two common ways to measure the error of floating-point computation: \emph{absolute round-off error} and \emph{relative round-off error}. Let $x$ be the ideal mathematical result of a floating-point computation and $x_\mathbf{f}$ be the actual result with round-off errors. 
The absolute round-off error $Err_{abs}(x_\mathbf{f},x)$ and the relative round-off error $Err_{rel}(x_\mathbf{f},x)$ can be defined as: 
$Err_{abs}(x_\mathbf{f},x)=\left|x-x_\mathbf{f}\right|$ and $Err_{rel}(x_\mathbf{f},x)=\left|\frac{x-x_\mathbf{f}}{x}\right|$. 

Let $\mbox{\sl op}$ be the operation and $[.]$ be the evaluation function, $\mbox{\sl op}_\mathbf{f}$ be the corresponding floating-point operation of an arithmetic operation $\mbox{\sl op}$ under a floating-point format $\mathbf{f}$. The relationship between the evaluation $[\mbox{\sl op}]$ without round-off errors and the evaluation $[\mbox{\sl op}_\mathbf{f}]$ with round-off errors can be described as $[op_\mathbf{f}] = [op]\cdot(1+e)+d$, where $e$ is the \emph{relative error variable} and $d$ is the \emph{absolute error variable}, bounded by $|e|\leq \epsilon$ and  $|d|\le \delta$. 

Here, we combine $\epsilon$ and $\delta$ to create a unified model that accommodates both normalized and denormalized floating-point numbers. The actual values for $\epsilon,\delta$ depend on the floating-point format $\mathbf{f}$. For example, if $\mathbf{f}$ is the 32-bit single-precision format, then we have $\epsilon = 2^{-23},\delta = 2^{-149}$ when considering arbitrary rounding mode, and $\epsilon = 2^{-24},\delta = 2^{-150}$ when considering nearest-to rounding mode.

\subsection{First-Order Differential Characterization (fo-DC)}\label{sect:fodc}

FPTaylor \cite{fptaylor} proposes to use symbolic Taylor expansions to compute tight bounds for the absolute round-off error of floating-point expressions. 
It abstracts the floating-point implementation of function $f(\mathbf{x})$ over a vector $\mathbf{x}$ of program variables into a real-valued function $\hat{f}(\mathbf{x}, \mathbf{e}, \mathbf{d})$,  where $\mathbf{e}=(e_1,\dots, e_k)$ and 
resp. $\mathbf{d}=(d_1,\dots,d_k)$ are vectors of relative and absolute error variables such that each $(e_i,d_i)$ corresponds uniquely to a floating-point operation in the computation of the function $f$. Hence $f(\mathbf{x})=\hat{f}(\mathbf{x},\mathbf{0},\mathbf{0})$. To compute a bound on the \emph{absolute round-off error} $\max_{\mathbf{x}\in B} |f(\mathbf{x}) - \hat{f}(\mathbf{x},\mathbf{e},\mathbf{d}))|$ of $f$ over a mandatory bounded region $B$, FPTaylor applies a first-order differential characterization (fo-DC)  to the abstracted function $\hat{f}(\mathbf{x},\mathbf{e},\mathbf{d})$ around the point $(\mathbf{x},\mathbf{0},\mathbf{0})$ (where $\mathbf{0}$ is the zero vector), and obtains $$
\textstyle \hat{f}(\mathbf{x}, \mathbf{e}, \mathbf{d}) = \hat{f}(\mathbf{x}, \mathbf{0}, \mathbf{0}) + \sum_{i=1}^{k} \frac{\partial \hat{f}}{\partial e_i}(\mathbf{x}, \mathbf{0}, \mathbf{0})\cdot e_i + \sum_{i=1}^{k} \frac{\partial \hat{f}}{\partial d_i}(\mathbf{x}, \mathbf{0}, \mathbf{0})\cdot d_i + R_2(\mathbf{x}, \mathbf{e}, \mathbf{d})
$$ and  
$$
\textstyle R_2(\mathbf{x}, \mathbf{e}, \mathbf{d}) = \frac{1}{2} \sum_{i,j=1}^{2k} \frac{\partial^2 \hat{f}}{\partial y_i \partial y_j}(\mathbf{x}, \mathbf{e}', \mathbf{d}') y_i y_j
$$
where $y_1,...,y_{2k}$ range over $e_1,\ldots,e_k$, $d_1,\ldots,d_k$, $\mathbf{e}'\in \mathbb{R}^{k}$ satisfies $|e'_i|\leq \epsilon$ and $\mathbf{d}'\in \mathbb{R}^{k}$ satisfies $|d'_i|\leq \delta$ for $i=1,\ldots,k$. Since $\hat{f}(\mathbf{x}, 0, 0) = f(\mathbf{x})$, one has: 

\begin{equation}\label{eq:fodc}
\small\textstyle\left| \hat{f}(\mathbf{x}, \mathbf{e}, \mathbf{d}) - f(\mathbf{x}) \right| \le   \sum_{i=1}^{k} \left| \frac{\partial \hat{f}}{\partial e_i} (\mathbf{x}, 0, 0) \right|\cdot \epsilon + \sum_{i=1}^{k} \left| \frac{\partial \hat{f}}{\partial d_i} (\mathbf{x}, 0, 0)\right|\cdot \delta + \left| R_2(\mathbf{x}, \mathbf{e}, \mathbf{d}) \right|
\end{equation}

To compute an upper bound $\gamma$ for the absolute round-off error $\textstyle|\hat{f}(\mathbf{x}, \mathbf{e}, \mathbf{d}) - f(\mathbf{x})|$, FPTaylor uses rigorous global optimization techniques to maximize the expressions $\textstyle| \frac{\partial \hat{f}}{\partial e_i} (\mathbf{x}, 0, 0) |, | \frac{\partial \hat{f}}{\partial d_i} (\mathbf{x}, 0, 0)|, | R_2(\mathbf{x}, \mathbf{e}, \mathbf{d}) |$ in (\ref{eq:fodc}) over the bounded region $B$, so that an upper bound $\gamma$ can be derived.

\subsection{Floating-Point Programs}\label{sec:fp}

Our floating-point programs have float and integer valued program variables and constants, as well as basic arithmetic operations namely addition, subtraction, multiplication, and division. Arithmetic operations incur round-off errors when either at least one of the involved program variable or constant is float or the operation is division, and do not incur round-off errors for the other cases. We treat integer/float overflow and division by zero as exceptions, and focus on program behaviors without such exceptions. We have standard program constructs such as conditional branching and while loops. The detailed syntax is as follows:
\begin{eqnarray*}\textstyle
P &::=& x := \alpha~\mid~ P_1;P_2~\mid~\textbf{if}~b~\textbf{then}~P_1~\textbf{else}~P_2~\mid~\textbf{while}~b~\textbf{do}~P \\
\alpha &::=& ~x~\mid~c~\mid~\alpha_1 + \alpha_2~\mid~\alpha_1 - \alpha_2~\mid~\alpha_1 * \alpha_2~\mid \alpha_1 / \alpha_2 \\ b &::=& \alpha_1\le \alpha_2~\mid~ \alpha_1 < \alpha_2~\mid~\neg b ~\mid~ b_1\wedge b_2~\mid~b_1\vee b_2   
\vspace{-0.1cm}
\end{eqnarray*}
In the above syntax, $x$ is a program variable and $c$ is a constant.
$\alpha$ is an arithmetic expression that involves addition, subtraction, multiplication, and division, 
and $P$ is a program that is built from program constructs including assignment statements ($x := \alpha$), sequential composition ($P_1;P_2$), conditional branches ($\textbf{if}~b~\textbf{then}~P_1~\textbf{else}~P_2$) and while loops ($\textbf{while}~b~\textbf{do}~P$). The meaning of these program constructs is standard. 

The semantics of our floating-point programs is given operationally via \emph{floating-point control flow graphs} to be defined below. Informally, a floating-point control flow graph describes how program executes under the perturbation of the round-off errors incurred from arithmetic operations.

\smallskip
\noindent\emph{floating-point CFGs.}  A \emph{floating-point control flow graph} (fp-CFG) is a tuple $(L, X, R, \mbox{\sl Init}, \rightarrow)$ where $L$ is a finite set of \emph{locations} with a designated \emph{initial} location $\ell^*$, 
$X$ is a finite set of \emph{program variables}, 
$R$ is a finite set of \emph{error variables} arising from some underlying floating-point model, 
$\mbox{\sl Init}$ is a formula over program variables that acts as the \emph{initial condition}, and 
$\rightarrow$ is a finite set of \emph{transitions}. 
Each transition is a tuple $(\ell, b, F, R', \ell')$ where $\ell$ is the \emph{source} location before taking the transition, $b$ is the \emph{guard condition} that serves as the condition to execute the transition and is a propositional formula whose atomic propositions are inequalities between arithmetic expressions of program variables, $R'\subseteq R$ is a subset of error variables that are relevant to the transition, $F$ is an \emph{update function} that takes a program valuation $\mathbf{v}$ (over $X$) with an error valuation $\mathbf{r}$ (over $R'$) and outputs a program valuation $F(\mathbf{v},\mathbf{r})$ (over $X$) that represents the program valuation after executing the transition, and $\ell'$ is the \emph{target} location after taking the transition.  

W.l.o.g, we assume that the error variables of different transitions are disjoint. We allow $R'=\emptyset$ in a transition $(\ell, b, F, R', \ell')$ so that the update function $F$ does not incur floating-point errors, which is useful for pure integer arithmetic that has no floating-point errors. Moreover, we assume that each guard condition $b$ is a conjunction of inequalities between arithmetic expressions of program variables. Note that a guard condition that may involve negation and disjunction can be first transformed into disjunctive normal form and then each disjunctive clause can be split into a standalone transition.

Informally, in an fp-CFG of a floating-point program, a location refers to a program counter or a cut point of the program. The set of program variables X consists of variables appearing in the program, and the set of error variables R corresponds to floating-point errors associated with the floating-point operations. A transition $(\ell, b, F, R', \ell')$ represents a (possibly multi-step) program fragment from location $\ell$ to $\ell'$, which may correspond to a basic block consisting of a sequence of assignments. Starting from location $\ell$ with a program valuation $\mathbf{v}$, if $\mathbf{v}$ satisfies the guard condition b up to floating-point errors, then the program may move to location $\ell'$, where the next valuation is updated to $F(\mathbf{v}, \mathbf{r})$ for some error valuation $\mathbf{r}$ over $R'$.

\subsection{Transformation from Programs into fp-CFG's}

Below we list the major aspects of transforming a floating-point control-flow graph (fp-CFG) into a program as follows.

\begin{itemize}
\item If a location $\ell$ corresponds to an assignment statement $x := \alpha$, then there is a transition $(\ell, \mathbf{true}, F, R', \ell')$, where $\ell'$ is the successor program counter after the assignment, $F$ is the update function derived from the assignment with round-off errors taken into account, and $R'$ is the set of error variables involved in $F$.

\item If a location $\ell$ corresponds to a conditional branch
$\textbf{if}~b~\textbf{then}~P_1~\textbf{else}~P_2$, then there are two transitions
$(\ell, b, \mbox{\sl id}, \emptyset, \ell_{\textbf{then}})$ and
$(\ell, \neg b, \mbox{\sl id}, \emptyset, \ell_{\textbf{else}})$.
Here, $\mbox{\sl id}$ denotes the identity update function (i.e., $\mbox{\sl id}(\mathbf{v}, \mathbf{r}) = \mathbf{v}$ for all $\mathbf{v}, \mathbf{r}$), and $\ell_{\textbf{then}}$ (resp.\ $\ell_{\textbf{else}}$) is the entry location of the \textbf{then} (resp.\ \textbf{else}) branch.

\item If a location $\ell$ is the entry point of a while loop with guard $b$, then it has two transitions
$(\ell, b, \mbox{\sl id}, \emptyset, \ell')$ and
$(\ell, \neg b, \mbox{\sl id}, \emptyset, \ell'')$,
where $\ell'$ is the program counter of the first statement in the loop body, and $\ell''$ is the successor program counter after the loop. Moreover, if a location $\ell$ corresponds to the last statement of the loop body, then its successor location is the loop entry point.
\end{itemize}

Alternatively, one may select specific cut points (e.g., conditional branches or loop entry points) and construct transitions between these cut points by summarizing straight-line code segments into single update functions.

An example for the transformation is given below.

\begin{figure}
  \begin{subfigure}[t]{0.49\textwidth}
    \centering
    \begin{lstlisting}
    #Example SineNewton
    #precondition: -1<=x<=1
    int i = 0;
    while( i < 10 ){   
        x = x - sin(x) / cos(x);
        i = i + 1;  
    }
    \end{lstlisting}
  \end{subfigure}%
  \begin{subfigure}[t]{0.49\textwidth}
    \centering
    \vspace{0pt}
    \includegraphics[width=\textwidth]{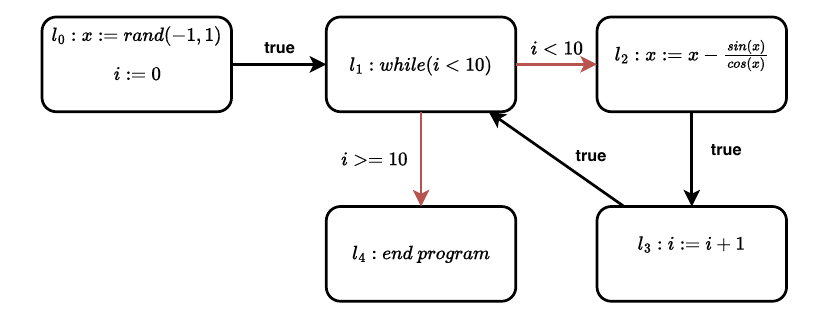}
  \end{subfigure}\hfill
  \caption{SineNewton (left) and its fp-CFG (right)}
  \label{fig:cfg}
\end{figure}
\begin{example}
Let's consider the following program, which uses Newton iteration to calculate  the root of $sin(x)=0$. For this program, we have the fp-CFG in Figure~\ref{fig:cfg}. $cos(x)$ and $sin(x)$ can be replaced with Taylor expansion and hence treated as simple polynomials about x. In this graph, we have a location for each statement in the program (including the termination location). Each edge indicates a transition whose label corresponds to the condition to execute the transition and whose content specifies its update function. To make the graph easy to understand, we omit the error variables since they are implicitly given in the expressions of the program. \qed
\end{example}

\subsection{Relaxation of Guard Condition $b$}

To give a formal semantics to fp-CFGs, special care must be taken with guard conditions, as a guard $b$ may involve floating-point computations and thus incur round-off errors. To account for this, we define the rounding-error interpretation $\langle b \rangle$ of a guard condition $b$ recursively based on its structure.

\begin{itemize}
\item If $b$ is an atomic inequality of the form $\alpha_1 \le \alpha_2$, then $\langle b \rangle$ is defined as the quantified formula
\[
\exists \mathbf{r}_1, \mathbf{r}_2.\;
(\alpha_1[\mathbf{r}_1] \le \alpha_2[\mathbf{r}_2])
\;\wedge\;
|\mathbf{r}_1| \le \kappa_1
\;\wedge\;
|\mathbf{r}_2| \le \kappa_2,
\]
where:
(i) each $\mathbf{r}_i$ ($i=1,2$) is a vector of fresh variables representing the relative or absolute error variables arising from the floating-point evaluation of $\alpha_i$;
(ii) each expression $\alpha_i[\mathbf{r}_i]$ is obtained by inserting the corresponding error variables into $\alpha_i$ according to the floating-point model; and
(iii) each bound $|\mathbf{r}_i| \le \kappa_i$ specifies coordinate-wise error bounds (i.e., $\epsilon$ and $\delta$ as defined in Section~\ref{sect:fpmodel}). Other atomic forms of guard conditions (e.g., $\alpha_1 < \alpha_2$) are handled analogously.

\item For the recursive case, if $b = b_1 \wedge b_2$, then we define
\[
\langle b_1 \wedge b_2 \rangle := \langle b_1 \rangle \wedge \langle b_2 \rangle.
\]
\end{itemize}

Given an fp-CFG $(L, X, R, \mbox{\sl Init}, \rightarrow)$, a \emph{program state} $\sigma$ is a pair $(\ell, \mathbf{v})$ where $\ell$ is a location that represents the current location of the fp-CFG, and $\mathbf{v}$ is a program valuation that specifies the current program values for every program variable in $X$. The execution of an fp-CFG is given by the notion of paths: a \emph{path} is a finite sequence of program states $(\ell_0, \mathbf{v}_0),\dots,(\ell_n, \mathbf{v}_n)$ such that 
\begin{itemize}
\item $(\ell_0, \mathbf{v}_0)$ is the \emph{initial} program state of the path that fulfills the formula $\mbox{\sl Init}$, i.e., $\ell_0=\ell^*$ and $\mathbf{v}_0\models\mbox{\sl Init}$, and 
\item For each $0\le j\le n-1$, there exists a transition $(\ell, b, F, R', \ell')$ such that $(\ell, \ell') = (\ell_j, \ell_{j+1})$, $\mathbf{v}\models \langle b\rangle$, and $\mathbf{v}_{j+1} = F(\mathbf{v}_j, \mathbf{r})$ for some error valuation $\mathbf{r}$ over $R'$. If $R'=\emptyset$, then we simply ignore the parameter $\mathbf{r}$ in $F$.  
\end{itemize}
It is worth noting that under floating-point perturbations, the satisfaction of the floating-point version $\langle b\rangle$ of a guard condition $b$ for a program valuation $\mathbf{v}$ can be \emph{unstable}, i.e., it is possible that both $\mathbf{v}\models \langle b\rangle$ and $\mathbf{v}\models \langle \neg b\rangle$ hold due to different round-off errors. Therefore, we take more than one path into account when considering invariants due to the non-determinism.

Below, we define invariants. The definition is presented directly over fp-CFGs. 
Informally, an invariant is a map that assigns to each location of an fp-CFG a formula that over-approximates the reachable program states to the location. 

\smallskip
\noindent\emph{Invariants.} Given an fp-CFG $(L, X, R,  \mbox{\sl Init}, \rightarrow)$, an \emph{assertion map} is a map $I$ that assigns to every location $\ell\in L$ 
an assertion $I(\ell)$ over the program variables $X$. An \emph{invariant} is an assertion map $I$  
such that for any path $(\ell_0, \mathbf{v}_0),\dots,(\ell_n, \mathbf{v}_n)$, we have that 
$\mathbf{v}_j\models I(\ell_j)$ for every $0\le j \le n$. Furthermore, an invariant $I$ is \emph{inductive} if it satisfies the following conditions:

\begin{itemize}
\item (\emph{Initiation}) The initial condition $\mbox{\sl Init}$ implies the invariant. Formally, for any program valuation $\mathbf{v}$, it holds that 
\vspace{-0.1cm}
\begin{equation}\label{eq:initiation}
\textstyle\mathbf{v}\models \mbox{\sl Init} \Rightarrow \mathbf{v}\models I(\ell^*). 
\end{equation}
\item (\emph{Consecution}) The invariant is preserved in every transition: for any transition $(\ell, b, F, R', \ell')$, for any program valuation $\mathbf{v}$ and error valuation $\mathbf{r}$ over $R'$, it holds that
\begin{equation}\label{eq:consecution}
\textstyle(\mathbf{v}\models (I(\ell) \wedge \langle b\rangle)) \Rightarrow F(\mathbf{v},\mathbf{r}) \models I(\ell').
\end{equation}
\end{itemize}

It is straightforward (by induction on the length of a path) to observe that an inductive invariant is an invariant. 

\section{Overview}

Consider the SineNewton example from FPbench~\cite{fpbench} below. 

\begin{lstlisting}
#precondition: -1<=x<=1
int i = 0;
float px, qx;

#loc1: invariant I(loc1)(x,i)
while( i < 10 ){ 
    px = x-(x^3)/6+(x^5)/120+(x^7)/5040;
    qx = 1-(x^2)/2+(x^4)/24+(x^6)/720;
    x = x - px/qx;
    i = i + 1;
}
#loc2: invariant I(loc2)(x,10)
#range [low, up] of x
\end{lstlisting} 

The example roughly models the computation of the root of $\sin(x)=0$ via Newton's method. It is similar to the SineNewton example in Section 2 but uses the Taylor series to replace the $\sin$ and $\cos$ functions\footnote{The assignment for $\mbox{\sl qx}$ is not exactly the Taylor expansion of $\cos$, but we keep the original benchmark from FPbench.}. In the example, $x$ is the key program variable that records the current value in the Newton's iteration, and $i$ is the loop counter. The loop executes 10 times of Newton's iteration and returns the final value of $x$. The precondition for the value of $x$ is given by $-1\le x \le 1$. Note that $x$ is a float variable that incurs rounding errors, while $i$ is an integer variable that does not incur rounding error.

Our goal is to automatically generate an invariant for this loop under the rounding errors incurred in the calculation of $\mbox{\sl px}, \mbox{\sl qx},\mbox{\sl x}$. The precision of the generated invariant is measured by the range $[\mbox{\sl low}, \mbox{\sl up}]$ of the final value of $x$ derivable from the invariant at loop termination. 

Below we illustrate the main steps of our approach for this example. We consider two cut points loc1, loc2 in the fp-CFG, for which the location loc1 is at the entry point of the loop (marked by "invariant I(loc1)(x,i)"), and the location loc2 is at the termination (marked by "invariant I(loc2)(x,10)").
The update function for the transition from loc1 to itself that executes the loop body once is denoted by $F$. 

Our goal is to solve a polynomial invariant assertion $I(\mbox{loc1})(x,i)$ at the location loc1, obtain the post condition at loc2 as $I(\mbox{loc1})(x,10)$ (since the value of the integer variable $i$ at termination is $10$), and optimize the invariant assertion $I(\mbox{loc1})$ so that the range $[\mbox{\sl low}, \mbox{\sl up}]$ of the value of $x$ at termination that is derivable from $I(\mbox{loc1})(x,10)$ (i.e., $I(\mbox{loc1})(x,10)\implies \mbox{\sl low}\le x\le \mbox{\sl up}$) is as small as possible. 

\smallskip
\noindent{\textbf{Step 1.}} The first step is to determine or guess a (tentative) initial invariant assertion $B(x,i)$ at the location loc1 that holds for every iteration of the loop. For this example, it is straightforward to observe that $0\le i\le 10$ holds for every loop iteration. Moreover, with the help of an SMT solver (e.g. Z3~\cite{z3}, Polyqent~\cite{poly2}), one can verify that \( -1 \leq x \leq 1 \) holds for every loop iteration without considering the rounding error. Therefore, we \emph{guess} that $B(x,i)=0\le i\le 10\wedge -1 \leq x \leq 1$. The guess is validated by the polynomial invariant generated by our approach in the last step.  

\smallskip
\noindent{\textbf{Step 2.}} Using the tentative initial invariant assertion $B(x,i)$, we determine an upper bound \( \gamma_F \) for the absolute rounding error of the update function $F$ (i.e., the absolute error for one loop iteration).
We use existing floating-point analyzers (such as FPTaylor\cite{fptaylor}) to achieve this. Given the range of program values specified by $B(x,i)$, we find an upper bound \( \gamma_F = 7.115443 \cdot 10^{-7} \) by FPTaylor. 

\smallskip
\noindent{\textbf{Step 3.}} The next step is to establish constraints for the invariant $I(\mbox{loc1})(x,i)$ to be resolved. Recall that the basic constraints for the invariant $I$ includes the initiation (\ref{eq:initiation}) and the consecution condition (\ref{eq:consecution}). We keep the initiation condition, and relax the consecution (\ref{eq:relaxation1}) by using $\gamma_F$ as an over-approximation for the rounding error of one loop iteration. The relaxed consecution condition is given as follows:
\[
\forall x,x',i. \left[\left(-1 \leq x \leq 1 \land 0 \leq i \leq 9 \land I(\mbox{loc1})(x,i) \land -\gamma_F \leq x' - \left( x - \frac{px}{qx} \right) \leq \gamma_F\right) \implies I(\mbox loc1)(x', i+1)\right]
\]
for which the fresh variable $x'$ represents the value of $x$ after one loop iteration, and $px, qx$ here are short hand for the polynomial expressions $x-x^3/6+x^5/120+x^7/5040$ and $1-x^2/2+x^4/24+x^6/720$ without rounding errors, respectively. Note that the use of the upper bound $\gamma_F$ soundly eliminates the error variables corresponding to the rounding errors in the calculation of $x - (px/qx)$. 
This is the key to reduce the burden of constraint solving and thus making constraint-solving methods for the invariant generation of floating-point programs possible.

\smallskip
\noindent{\textbf{Step 4.}} We solve the invariant $I(\mbox{loc1})(x,i)$ by the template-based method~\cite{DBLP:conf/pldi/AsadiC0GM21, DBLP:conf/cav/ColonSS03,DBLP:conf/sas/SankaranarayananSM04,polyqent,DBLP:journals/pacmpl/LiuFYSL22,DBLP:journals/toplas/WuWXZZY25} that first assigns a polynomial template with unknown coefficients and then solves the unknown coefficients in the template w.r.t the constraints established from the previous step to obtain a concrete polynomial invariant. We adopt existing solving methods such as Handelman's Theorem (Theorem~\ref{thm:handelman}) for the resolution of the unknown coefficients. Moreover, we propose the following novelties in the solving of the invariant-relevant constraints: a constraint simplification technique via barrier certificate and optimization methods to minimize the range $[\mbox{\sl low},\mbox{\sl up}]$ for the program variable $x$ at termination. 
For the optimization, we propose two options: the first is simply to set up the objective function $\mbox{\sl minimize}(\mbox{\sl up} - \mbox{\sl low})$, while the second is to perform binary search over possible values of $\mbox{\sl up},\mbox{\sl low}$. 

For this example, with the objective function $\mbox{\sl minimize}(\mbox{\sl up} - \mbox{\sl low})$ and applying (the sound form of) Handelman's Theorem, our approach is able to solve a polynomial invariant $I(\mbox{loc1})(x,i)$ that achieves \( \mbox{\sl low} = -0.06 \) and \( \mbox{\sl up} = 0.06 \). This shows that after the execution of the loop, the value of the variable $x$ falls in the interval $[-0.06, 0.06]$. We are also able to  verify the initial guess of the tentative invariant assertion $B(x,i)$ by checking the validity of the implication
\[
I(\mbox{loc1})(x,i) \land 0 \leq i \leq 9 \implies -1\leq x\leq 1.
\]
Note that we only need to verify the situation within the loop (i.e., $0 \leq i \leq 9$), since the situation at loop termination (i.e., $i=10$) is already entailed by the values of $\mbox{\sl low}$ and $\mbox{\sl up}$. Therefore, we obtain a polynomial invariant $I(\mbox{loc1})(x,i)$ that minimizes the range $[\mbox{\sl low},\mbox{\sl up}]$ of the variable $x$ (up to algorithmic parameters) at termination. In addition, we use coarse bounds on the variables to estimate the maximum error via an external solver. In this case, the maximum error is $1.7170882500149726 \times 10^{-16}$, which is well within the tolerance handled by the numerical repair technique introduced in Section~\ref{sec:NR}. Therefore, the result is confirmed to be valid.

\section{An Framework for Bounded floating-point Programs}

In this section, we propose a polynomial invariant generation framework for floating-point programs whose variables take bounded values. Bounded programs 
are of practical importance, as variables in most real-world applications have bounded values. Unbounded programs usually cause unbounded round-off errors and overflow. 

\subsection{Details in Our Framework}
\label{sec:DA}
\setcounter{secnumdepth}{3}

Let $\Gamma=(L, X, R, \mbox{\sl Init}, \rightarrow)$ be an fp-CFG of the input program, and $d$ be the input polynomial degree. Our framework contains four stages to generate a polynomial invariant of degree $d$ for the fp-CFG $\Gamma$.

\subsubsection{Preparation Stage} 

To obtain bounds on floating-point errors, the framework first chooses an assertion map $B$ that gives coarse bounded ranges $\{\mathbf{v}\mid \mathbf{v}\models B(\ell)\}$ for all locations $\ell$. 
At this stage, we do not know whether $B$ is valid (i.e., an invariant) or not.
The only requirement is that $Init \models B$, where $Init$ denotes the initial condition of the fp-CFG. 
The guess of $B$ can be done via 
multiple program simulations or 
other invariant generation techniques, such as 
~\cite{tvpifp,pine}.

Given the coarse ranges in $B$, 
we apply round-off error analysis (e.g.,~\cite{fptaylor,MagronCD17,DarulovaINRBB18}) to each update function $F$ emanating from a location $\ell$ with the range $B(\ell)$ as the input range. This yields an upper-bound vector $\gamma_F$, each component of which bounds the round-off error of a distinct program variable after the update assignment of the function, such that
\(\textstyle
\left| F(\mathbf{x}, \mathbf{r}) - F(\mathbf{x}, \mathbf{0}) \right| \le \gamma_F
\)
for all $\mathbf{x} \models B(\ell)$ and all error valuations $\mathbf{r}$. In our current implementation, we apply the fo-DC method~(\ref{eq:fodc}) from FPTaylor~\cite{fptaylor}, as described in Section~\ref{sect:fodc}. In this way, we eliminate the error variables involved in the update function and retain only the upper-bound vector $\gamma_F$ for the incurred round-off errors. 

Then, given the input degree $d$, we construct a polynomial template $\eta$. Specifically, for each location $\ell$, the template is defined as
\(\textstyle
\eta_\ell(\mathbf{x}) = \sum_{i} c_{\ell,i} \cdot q_i(\mathbf{x}),
\)
where $q_i(\mathbf{x})$ enumerates all monomials formed by finite products of program variables with degree at most $d$, and $c_{\ell,i}$ are unknown coefficients. 

\subsubsection{Constraint Deriving Stage}

Our target invariant is of the form $I \equiv \eta \ge 0$. 
To ensure that $I$ is an inductive invariant, we impose 
the initiation condition~(\ref{eq:initiation}) and the consecution condition~(\ref{eq:consecution}) for all transitions in the fp-CFG.
To adapt these constraints in the presence of floating-point errors, we first over-approximate the guard conditions, and then strengthen the initiation and consecution constraints. 

\smallskip
\noindent\textbf{Relaxation of guard conditions.}
We replace each guard condition $b$ in a transition emanating from a location $\ell$ with an over-approximation $\overline{b}$ that accounts for round-off errors similar to~\cite{titolo2018}. The formula $\overline{b}$ is defined recursively as follows:

\begin{enumerate}
\item \emph{Base case ($\le, <$).}  
When $b$ is $\alpha_1 \le \alpha_2$, let $F_i$ ($i=1,2$) be the function that maps a program valuation $\mathbf{v}$ to the evaluation of the arithmetic expression $\alpha_i$ under $\mathbf{v}$ without round-off errors. Given the input range $B(\ell)$ from the preparation stage, 
we derive a scalar bound $\gamma_{F_i}$ on the absolute round-off error of $F_i$ using the fo-DC method, and define the over-approximation
\(\textstyle
\overline{b} := \alpha_1 - \gamma_{F_1} \le \alpha_2 + \gamma_{F_2}.
\) For $b=\alpha_1 < \alpha_2$, we relax it to $\alpha_1 \le \alpha_2$ and then perform the same over-approximation. 
\item \emph{Recursive case.}  
When $b = b_1 \wedge b_2$, 
$\overline{b}$ is defined as $\overline{b} := \overline{b}_1 \wedge \overline{b}_2$.
\end{enumerate}
Note that $\gamma_{F_i}$ above bounds the round-off errors when $B$ provides valid ranges, so we have that $\langle b \rangle$ implies $\overline{b}$ when $B$ is a valid invariant.

\smallskip
\noindent\textbf{Strengthening of consecution conditions.}
Given a transition $(\ell, b, F, R', \ell')$, we have two ways to strengthen the consecution condition w.r.t round-off errors.

\smallskip
\noindent{\em First strengthening.}
We introduce a fresh variable $x'$ for each program variable $x$ in the fp-CFG. Let $\mathbf{x}=(x_1,\dots,x_{|X|})^{\mathrm{T}}$ denote the vector of program variables and $\mathbf{x}'=(x'_1,\dots,x'_{|X|})^{\mathrm{T}}$ the vector of primed variables, representing the values of the program variables after one transition. Let $\mathbf{0}=(0,\dots,0)^{\mathrm{T}}$ denote the zero vector. Using the previously computed bounds $\gamma_F$ for the update function $F$, we strengthen the original consecution condition~(\ref{eq:consecution}) as follows:
\begin{equation}\label{eq:relaxation1}
\textstyle\small\forall \mathbf{x}, \mathbf{x}'.\;
\left[
\left(
(\mathbf{x} \models I(\ell) \wedge B(\ell) \wedge \overline{b})
\;\wedge\;
-\gamma_F \le \mathbf{x}' - F(\mathbf{x},\mathbf{0}) \le \gamma_F
\right)
\Rightarrow
\mathbf{x}' \models I(\ell')
\right].
\end{equation}

The intuition is that we replace the floating-point guard $\langle b \rangle$ with its over-approximation $\overline{b}$ and use $\gamma_F$ to bound the deviation between the perturbed values $\mathbf{x}'$ 
and the non-perturbed values $F(\mathbf{x},\mathbf{0})$. As a result, error variables from the computation of $F$ are eliminated from subsequent constraint solving, potentially reducing the computational cost significantly. If the guessed assertion map $B$ is a valid invariant, then the strengthened condition~(\ref{eq:relaxation1}) implies, and thus is a sound under-approximation of, the original consecution condition~(\ref{eq:consecution}). The strengthened consecution conditions for all transitions are combined together conjunctively.

An advantage of this strengthening is that the right-hand side of the implication, $\mathbf{x}' \models I(\ell')$, remains simple. However, the left-hand-side constraint
$-\gamma_F \le \mathbf{x}' - F(\mathbf{x},\mathbf{0}) \le \gamma_F$
can be complex when $F$ contains high-degree polynomials or rational expressions, which can make invariant solving difficult. To address this issue, we propose an alternative strengthening with a simpler left-hand-side constraint and a more complex right-hand side constraint.

\smallskip
\noindent{\em Second strengthening.} For each program variable $x$, we introduce a fresh variable $r_x$ that represents the deviation between the perturbed value (due to round-off errors) and the non-perturbed value of $x$ after one transition. We collect these variables into a vector
$\mathbf{r}_X = (r_{x_1}, \dots, r_{x_{|X|}})^{\mathrm{T}}$.
Using these variables, we strengthen the consecution condition~(\ref{eq:consecution}) as follows:
\begin{equation}\label{eq:relaxation2}
\textstyle\small\forall \mathbf{x}, \mathbf{r}_X.\;
\left[
\left(
(\mathbf{x} \models I(\ell) \wedge B(\ell) \wedge \overline{b})
\;\wedge\;
-\gamma_F \le \mathbf{r}_X \le \gamma_F
\right)
\Rightarrow
F(\mathbf{x},\mathbf{0}) + \mathbf{r}_X \models I(\ell')
\right].
\end{equation}
Here, $F(\mathbf{x},\mathbf{0}) + \mathbf{r}_X$ represents the next program state, and the deviation vector $\mathbf{r}_X$ is bounded by $\gamma_F$. As before, the floating-point guard $\langle b \rangle$ is replaced by $\overline{b}$. Again, if the guessed assertion map $B$ is an invariant, then the strengthened condition~(\ref{eq:relaxation2}) implies 
the original consecution condition~(\ref{eq:consecution}).

\smallskip
\noindent{\em Rewrite constraints.} We now rewrite the constraints into a form suitable for solving the unknown coefficients. This is done by replacing each occurrence of the target invariant $I$ with the polynomial template $\eta$. As a result, the constraints become a collection of quantified formulas, each of which takes one of the following forms.

For the initiation condition, the corresponding constraint has the form
\begin{equation}\label{eq:inittemplate}\textstyle
\forall \mathbf{x}.\, \left[\bigwedge_i h_i(\mathbf{x}) \ge 0 \;\Rightarrow\; \eta_{\ell^*}(\mathbf{x}) \ge 0\right].
\end{equation}
where each $h_i(\mathbf{x})$ is an arithmetic expression without unknown coefficients from the template $\eta$. 
For a strengthened consecution condition in (\ref{eq:relaxation1}), 
the corresponding constraint has the form
\begin{equation}\label{eq:extracted1}\textstyle
\forall \mathbf{x}, \mathbf{x}'.\, \left[
(\eta_{\ell}(\mathbf{x}) \ge 0
\;\wedge\;
\bigwedge_i h_i(\mathbf{x}, \mathbf{x}') \ge 0)
\;\Rightarrow\;
\eta_{\ell'}(\mathbf{x}') \ge 0
\right],
\end{equation}
where (a) $\mathbf{x}'$ is the vector of primed program variables representing the program variables after one transition, and (b) each $h_i(\mathbf{x}, \mathbf{x}')$ is an arithmetic expression that does not contain the unknown coefficients. 
For a strengthened consecution condition in (\ref{eq:relaxation2}), the corresponding constraint has the form
\begin{equation}\label{eq:extracted2}\textstyle
\forall \mathbf{x}, \mathbf{r}_X.\, \left[
(\eta_{\ell}(\mathbf{x}) \ge 0
\;\wedge\;
\bigwedge_i h_i(\mathbf{x}, \mathbf{r}_X) \ge 0)
\;\Rightarrow\;
\eta_{\ell'}\big(F(\mathbf{x}, \mathbf{0}) + \mathbf{r}_X\big) \ge 0
\right],
\end{equation}
where $\mathbf{r}_X$ denotes the vector of deviation variables and each $h_i(\mathbf{x}, \mathbf{r}_X)$ is an arithmetic expression that does not contain the unknown coefficients. 

\smallskip
\noindent\textbf{Further strengthening.}
We observe that the template $\eta$ appears on both sides of the implication in the consecution constraints. This increases the difficulty of constraint solving, as it leads to general non-linear programming problems. To address this issue, we strengthen these constraints using barrier certificate conditions~\cite{DBLP:conf/cav/WangCXZK21,DBLP:conf/fm/WuFGWXZ24}. 
The main idea is to examine the change in the value of $\eta$ before and after a transition and to enforce that the value after the transition does not decrease relative to the value before the transition with a barrier constant. In addition, we introduce a nonnegative constant $t$ to further strengthen the constraints; this constant is later used during the validation stage. Formally, given a barrier constant $\mbox{\sl bar} > 0$ and $t\geq0$, we strengthen the constraints from the previous step as follows.

\begin{itemize}
\item For each constraint of the form~(\ref{eq:extracted1}), we strengthen it to
\begin{equation}\label{eq:simplified1}\textstyle
\forall \mathbf{x}, \mathbf{x}'.\, \left[
\bigwedge_i h_i(\mathbf{x}, \mathbf{x}') \ge 0
\;\Rightarrow\;
\eta_{\ell'}(\mathbf{x}') - \mbox{\sl bar} \cdot \eta_{\ell}(\mathbf{x}) \ge t
\right].
\end{equation}

\item For each constraint of the form~(\ref{eq:extracted2}), we strengthen it to
\begin{equation}\label{eq:simplified2}\textstyle
\forall \mathbf{x}, \mathbf{r}_X.\, \left[
\bigwedge_i h_i(\mathbf{x}, \mathbf{r}_X) \ge 0
\;\Rightarrow\;
\eta_{\ell'}\big(F(\mathbf{x}, \mathbf{0}) + \mathbf{r}_X\big)
- \mbox{\sl bar} \cdot \eta_{\ell}(\mathbf{x}) \ge t
\right].
\end{equation}
\end{itemize}
The initiation condition~(\ref{eq:inittemplate}) is further strengthened only with $t$: we strengthen $\eta_\ell (x)\ge 0$ to $\eta_\ell (x)\ge t$ for later numerical repair.

\subsubsection{Coefficient Solving Stage}

In this stage, we solve for the unknown coefficients in the template $\eta$ to obtain a concrete polynomial invariant $\ell \mapsto \eta_\ell(\mathbf{x}) \ge 0$. The constraints derived in the previous step consist of the initiation condition~(\ref{eq:inittemplate}) with strengthening and the strengthened consecution constraints~(\ref{eq:simplified1}) or~(\ref{eq:simplified2}). We assume that each polynomial $h_i$ appearing in these constraints contains no division. If division occurs, we rewrite $h_i = P/Q$ and replace the inequality $h_i \ge 0$ with $P\cdot Q \ge 0$ (assuming no division-by-zero). If the sign of $Q$ is known, this condition can be further simplified.

To eliminate program variables and obtain constraints that involve the unknown coefficients only, we apply classical positivity results, namely Handelman’s theorem and Putinar’s Positivstellensatz which is introduced in Section~\ref{sec:CE}. 

Now we get the equation system, solving it can be done by any external solver, but it typically admits infinitely many solutions. To solve a tight invariant, we introduce an optimization objective. Given a target program variable $x$, we introduce fresh variables $\mbox{\sl low}$ and $\mbox{\sl up}$ representing its lower and upper bounds, and require that the invariant implies
\begin{equation}\label{eq:optcst}
\forall \mathbf{v}.\; \left[\eta(\mathbf{v}) \ge 0 \Rightarrow \mbox{\sl low} \le \mathbf{v}(x) \le \mbox{\sl up}\right].
\end{equation}
We soundly under-approximate this condition by introducing a 
constant $a>0$ and enforcing
\begin{equation}\label{eq:optrelaxed}\textstyle
\left[x - \mbox{\sl up} \ge 0 \Rightarrow -\eta - a \ge 0\right]
\;\wedge\;
\left[-x + \mbox{\sl low} \ge 0 \Rightarrow -\eta - a \ge 0\right].
\end{equation}
Minimizing $\mbox{\sl up}-\mbox{\sl low}$ under these constraints yields a polynomial invariant that implies the tightest bound on the target variable $x$ with respect to the chosen objective. If the optimization problem cannot be solved by the external solver, we can apply a divide-and-conquer solution. Specifically, we perform a binary search over candidate values of $\mbox{\sl up}$ and $\mbox{\sl low}$ for each target variable within a given initial interval. This initial interval can be chosen as the range of the variable obtained from the initial assertion map $B$. For each candidate pair of values, we substitute concrete values for $\mbox{\sl up}$ and $\mbox{\sl low}$ and solve the resulting invariant constraints. Since $\mbox{\sl up}$ and $\mbox{\sl low}$ are fixed during each binary search iteration, the constraint system $\mathcal{C}_{\mathrm{handelman}}$ remains a linear program. Moreover, 
the under-approximation in~(\ref{eq:optrelaxed}) conform to the form required by~(\ref{eq:handelman}), allowing Handelman’s theorem (and similarly Putinar’s Positivstellensatz) to be applied directly. The second way would take a longer time, but as it gives less pressure to the external solver, it may find a tighter invariant.

\subsubsection{Validation Stage}

After solving the target invariant $I$, we must validate two sources that may affect soundness. The first concerns the guessed ranges in $B$. We check whether $I(\ell) \models B(\ell)$ for every location $\ell$. That is, whether $I$ implies $B$.
If so, then one can show that $B$ is a valid invariant, and the round-off bounds and the over-approximation of guards are also sound (see Section~\ref{sect:soundness} below). 
Otherwise, the guessed ranges in $B$ are insufficient; in this case, we enlarge the ranges in $B$ and repeat the entire framework starting from the first stage.

The second validation concerns numerical errors incurred by the external solver. Since the solver may not return exact solutions, we have an extra strengthening constant $t$ for all constraints, and solve the invariants with the extra constant $t$. Moreover, we accumulate the numerical errors associated with each unknown coefficient and compute an upper bound on their effect over the guessed ranges in $B$. We then compare this upper bound with the constant $t$. If the upper bound is smaller than the constant, then the error is covered and does not affect soundness. Otherwise, we report failure, indicating that the computed invariant may be unsound. This part is detailed in Section~\ref{sec:NR}.

The whole framework is summarized in Algorithm~\ref{alg:FloatInvGen-A}. A detailed walkthrough on the execution of the framework is given in the next part of the subsection.

\subsection{Methology}

\subsubsection{Theorem for Coefficient Solving}\label{sec:CE}

To eliminate program variables and obtain constraints solely over the unknown coefficients, we employ classical positivity results, in particular Handelman’s theorem and Putinar’s Positivstellensatz. We begin by presenting Handelman’s theorem, followed by Putinar’s Positivstellensatz.

\paragraph{Handelman’s Theorem.}

\begin{theorem}[Handelman~\cite{sheiderer}]\label{thm:handelman}
Let $\Theta=\{h_1,\dots, h_k\}$ be a finite set of linear (i.e., degree $1$) polynomials and $p$ be a polynomial. Suppose that the set $S:=\{\mathbf{v}\mid \forall 1\le i\le k. h_i(\mathbf{v})\ge 0\}$ is bounded. 
Then we have that $\forall \mathbf{v}. (\mathbf{v}\in S\Rightarrow p(\mathbf{v})\ge0)$ implies $p$ can be expressed as a finite sum $\sum_i \lambda_i\cdot g_i$ where each $g_i$ is a finite product (possibly with zero or duplicates) of polynomials from $\Theta$ and each $\lambda_i$ is a non-negative real coefficient. 
\end{theorem}

Building on the previous stage, each constraint has the form
\begin{equation}\label{eq:handelman}\textstyle
\forall \mathbf{v}.\; \left[(\bigwedge_{i=1}^k h_i(\mathbf{v}) \ge 0) \Rightarrow p(\mathbf{v}) \ge 0\right],
\end{equation}
where the polynomials $h_i$ contain no unknown coefficients and the coefficients of the polynomial $p$ is linear in the unknown template coefficients. By applying the reverse direction of Handelman’s theorem, we enforce the non-negativity of $p$ over the domain defined by $h_i \ge 0$ by expressing $p$ as
\begin{equation}\label{eq:handelmansum}\textstyle
p = \sum_{j=1}^{s_m} \lambda_j \cdot g_j, \qquad \lambda_j \ge 0,
\end{equation}
where $g_j$ ranges over all $s_m$ products of at most $m$ polynomials from $\{h_1,\dots,h_k\}$. By matching coefficients on both sides, we obtain a system of linear constraints over the unknown template coefficients and the fresh variables $\lambda_j$. Collecting such systems for all constraints yields a final linear constraint system $\mathcal{C}_{\mathrm{handelman}}$, whose solution determines the invariant.

\begin{proposition}
Let $m\ge 1$ be any upper bound on the number of polynomials in the finite summation form (\ref{eq:handelmansum}) of Handelman's Theorem, and $\mathcal{C}_{\mathrm{handelman}}$ be the final system of constraints under this upper bound as described previously. For all values $\mbox{\sl up}_1>\mbox{\sl up}_2$ for the special variable $\mbox{\sl up}$, if the constraint system $\mathcal{C}_{\mathrm{handelman}}$ with $\mbox{\sl up}=\mbox{\sl up}_2$ is feasible, then the same constraint system with $\mbox{\sl up}=\mbox{\sl up}_1$ is also feasible. Analogously, for any values $\mbox{\sl low}_1<\mbox{\sl low}_2$ for the variable $\mbox{\sl low}$, if the constraint system $\mathcal{C}_{\mathrm{handelman}}$ with $\mbox{\sl low}=\mbox{\sl low}_2$ is feasible, then the same constraint system with $\mbox{\sl low}=\mbox{\sl low}_1$ is also feasible.  
\end{proposition}

\begin{proof}
Let the target program variable for optimization be \( x \). If we have values \( \mbox{\sl up}_1, \mbox{\sl up}_2 \) such that \( \mbox{\sl up}_1 > \mbox{\sl up}_2 \), then \( x - \mbox{\sl up}_2 = (x - \mbox{\sl up}_1) + (\mbox{\sl up}_1 - \mbox{\sl up}_2) \). Thus, the result follows from the observation that \( x - \mbox{\sl up}_2 \) can be represented by \( x - \mbox{\sl up}_1 \) in the finite sum form under the same upper bound on the number of polynomials in a finite product. The case for the variable \( \mbox{\sl low} \) is similar. 
\end{proof}

\paragraph{Putinar’s Positivstellensatz.}

We now turn to Putinar’s Positivstellensatz. To this end, we first introduce the notion of \emph{sums of squares}.

\begin{definition}[Sums of Squares]\label{def:sos}
A polynomial \( p \) is a \emph{sum of squares} if we have that \( p = q_1^2 + \dots + q_m^2 \) for some polynomials \( q_1, \dots, q_m \). 
\end{definition}

In this work, we use a weakened form of Putinar's Positivstellensatz. The complete version involves the Archimedean condition; see, e.g.,~\cite{sheiderer}.

\begin{theorem}[Putinar~\cite{sheiderer}]\label{thm:put}
Let \( \Theta = \{h_1, \dots, h_k\} \) be a finite set of polynomials and \( p \) be a polynomial. Let \( S = \{\mathbf{v} \mid \forall 1 \le i \le k. h_i(\mathbf{v}) \ge 0\} \). Suppose that there exists some \( h_i \) such that the set \( \{\mathbf{v} \mid h_i(\mathbf{v}) \ge 0\} \) is compact. Then we have that \( \forall \mathbf{v}. (\mathbf{v} \in S \Rightarrow p(\mathbf{v}) > 0) \) implies that \( p \) can be expressed as a finite sum \( s_0 + \sum_{i=1}^k s_i \cdot h_i \) where each \( s_i \) is a sum of squares.  
\end{theorem}

To apply Putinar's Positivstellensatz, we again consider its reverse direction, using a finite sum \( \sum_i s_i \cdot h_i \) to witness the non-negativity of a polynomial \( p \) over a set \( S \). Moreover, we exploit the close connection between sums of squares and positive semidefinite matrices. In particular, for each constraint \(\mbox{\sl cst}\) in the form of (\ref{eq:handelman}) obtained from the previous steps, we establish the equality (Putinar form) 
\begin{equation}\label{eq:putinarform}
p = s_0 + \sum_{i=1}^k s_i \cdot h_i 
\end{equation}
where each \( s_i \) is written as \( \mathbf{g}^\mathrm{T} P_i \mathbf{g} \), with \( P_i \) being an unknown positive semidefinite matrix and \( \mathbf{g} \) the vector of monomials over program variables of degree at most a user-provided parameter \( m \). By matching coefficients on both sides of the above equality, we obtain the final constraints over the unknown template coefficients together with the unknown positive semidefinite matrices. Since the polynomials \( h_i \) do not involve any unknown template coefficients, the resulting constraint system is semidefinite and can be solved using semidefinite programming; see, e.g.,~\cite{polyqent}.

Finally, we justify the use of binary search under the Putinar formulation.

\begin{proposition}
Let \( m \ge 1 \) be any upper bound for the dimension of the positive semidefinite matrices in the sum-of-squares form (\ref{def:sos}), and \( \mathcal{C}_{\mathrm{putinar}} \) be the final system of constraints under this upper bound as described previously. For all values \( \mbox{\sl up}_1 > \mbox{\sl up}_2 \) for the special variable \( \mbox{\sl up} \), if the constraint system \( \mathcal{C}_{\mathrm{putinar}} \) with \( \mbox{\sl up} = \mbox{\sl up}_2 \) is feasible, then the same constraint system with \( \mbox{\sl up} = \mbox{\sl up}_1 \) is also feasible. Analogously, for any values \( \mbox{\sl low}_1 < \mbox{\sl low}_2 \) for the variable \( \mbox{\sl low} \), if the constraint system \( \mathcal{C}_{\mathrm{putinar}} \) with \( \mbox{\sl low} = \mbox{\sl low}_2 \) is feasible, then the same constraint system with \( \mbox{\sl low} = \mbox{\sl low}_1 \) is also feasible.   
\end{proposition}

\begin{proof}
The proof follows the same arguments, with the Handelman form~(\ref{eq:handelman}) replaced by the Putinar form~(\ref{eq:putinarform}).
\end{proof}

\subsubsection{Numerical Reparation}\label{sec:NR}

In our algorithm, we rely on an external solver to solve the resulting system of equations and obtain the invariant from the solved coefficients. However, due to various factors -- most notably floating-point arithmetic -- external solvers typically do not return exact solutions. Instead, they allow small numerical errors.

We illustrate this using the external solver Gurobi~\cite{gurobi}, which is used in our implementation. Gurobi requires a tolerance parameter that specifies the acceptable numerical error in constraint satisfaction. In particular, a constraint of the form $a \cdot x \le b$ is considered satisfied if
\[
a \cdot x - b \le \textit{TOL},
\]
where $\textit{TOL}$ denotes the solver’s tolerance value.

Since numerical errors from external solvers cannot be completely avoided, our goal is to ensure that they do not affect the soundness of the generated invariant. In our algorithm, in addition to using upper bounds on floating-point round-off errors and a small positive barrier constant, we introduce an additional small positive constant $t$. We strengthen the constraint
\[
\eta_{\ell'}(\mathbf{x}') - \mbox{\sl bar} \cdot \eta_{\ell}(\mathbf{x}) \ge 0
\]
to
\[
\eta_{\ell'}(\mathbf{x}') - \mbox{\sl bar} \cdot \eta_{\ell}(\mathbf{x}) \ge t,
\]
and use this margin $t$ to absorb numerical errors introduced by the external solver. Initiation condition \ref{eq:inittemplate} may also generate numerical error via external solver, hence should also be strengthened into  $\eta_{\ell}(x)\ge t$.

According to Handelman’s theorem~\ref{thm:handelman}, we construct an equation system by matching coefficients on both sides of Equation~(\ref{eq:handelmansum}). After solving this system with an external solver, we substitute the obtained solution back into the equations to estimate the numerical error in each coefficient of the template $\eta$. In the case of strengthened consecution condition, this yields an inequality
\[
\eta_{\ell'} - bar\cdot \eta_\ell -t = H + E
\]
where $ H$ is in the form of the right-hand side of (\ref{eq:handelmansum}), and $E$ is a polynomial that records the difference of coefficients between the two sides of the equality (of the form (\ref{eq:handelmansum})) due to numerical errors of external solver. The case for initiation condition is simpler:

\[\eta_{\ell^*} -t = H + E\]


To compute an upper bound for \(E\), we first need bounds for all variables appearing in \(E\), including both the program variables \(\mathbf{x}\) and the updated variables \(\mathbf{x}'\). Using the inferred invariant \(\eta_\ell(\mathbf{x}) \ge 0\), we first derive bounds for the program variables \(\mathbf{x}\). For the updated variables \(\mathbf{x}'\), we compute their ranges by evaluating the update function \(F(\mathbf{x},0)\) over the inferred invariant ranges and subsequently enlarging the resulting bounds using the error bound \(\gamma_F\) of the related updated functions. Based on these ranges, we then derive an upper bound \(\upsilon\) on the aggregate error term \(E\).
We then compare the upper bound $\upsilon$
with the additional strengthening constant $t$. If $\upsilon<t$, 
then the aggregated numerical error within $B(\ell)$ is covered by the strengthening, and the invariant remains valid despite solver inaccuracies. Otherwise, we return \textbf{False}, indicating that the soundness of the invariant cannot be guaranteed in this case.

The above numerical reparation procedure extends directly to the case of Putinar’s Positivstellensatz~\ref{thm:put}. The resulting sum-of-squares decomposition, obtained via semidefinite programming, may also incur numerical errors, which are handled analogously by introducing the strengthening margin $t$ and bounding the induced error term.

\paragraph{Example.}
Consider the constraint
\[
-1 \le x \le 1 \ \implies\ c_0 + c_1 x^2 \ge 0,
\]
and we guessed the variable $x$ with ranges over $[-1,1]$.

By Handelman’s theorem~\ref{thm:handelman}, we construct an equation system of the form
\[
h_1(1+x) + h_2(1-x) + h_3(1-x)^2 + h_4(1+x)^2 + h_5(1-x)(1+x)
= c_0 + c_1 x^2.
\]
Suppose the external solver returns values for $h_1,\dots,h_5,c_0,c_1$ with small numerical errors. Substituting the solution back, we no longer have exact equality, i.e.,
\[
c_0 + c_1 x^2 \ne h_1(1+x) + h_2(1-x) + h_3(1-x)^2 + h_4(1+x)^2 + h_5(1-x)(1+x).
\]
Let
\[
E(x) =  \big(h_1(1+x) + h_2(1-x) + h_3(1-x)^2 + h_4(1+x)^2 + h_5(1-x)(1+x)\big)-(c_0 + c_1 x^2)
\]
denote the numerical error. Let $\varepsilon = \sup_{x \in [-1,1]} |E(x)|$ be an upper bound on this error.

Thus, instead of guaranteeing
\[
c_0 + c_1 x^2 \ge 0,
\]
we only obtain
\[
c_0 + c_1 x^2 \ge -\varepsilon,
\]
which may violate the desired constraint.

Now suppose we strengthen the constraint to
\[
-1 \le x \le 1 \ \implies\ c_0 + c_1 x^2 \ge t,
\]
for some small constant $t > 0$. Using the same definition of $E(x)$ and same bound $\varepsilon$, we obtain
\[
c_0 + c_1 x^2 \ge t - \varepsilon.
\]
As long as $\varepsilon < t$, the right-hand side remains non-negative, ensuring that the invariant is preserved despite numerical errors.

\subsection{A Walkthrough of Our Framework}

Consider Example \texttt{ex1}~\cite{pine} below. This example consists of an infinite loop that updates the program variable $x$ using an affine expression. Therefore, we assign a single location $\ell$ at the entry point of the loop. In the corresponding fp-CFG, there is only one transition from $\ell$ to itself. The precondition specifies that the initial value of $x$ is zero and that the initial value of $i$ lies in the interval $[-1,1]$. Both $x$ and $i$ are floating-point variables. Let $F$ denote the update function of the loop body, given by the assignment
\(\textstyle
x = 1.5 \cdot x - 0.7 \cdot x + 1.6 \cdot i .
\)

    \begin{lstlisting}
#Example ex1
#precondition: -1<i<1 and x=0
while (true) {
    #invariant I(l)(x,i)
    x = 1.5 * x - 0.7 * x + 1.6 * i;  
}
\end{lstlisting} 

\noindent{\em Preparation Stage.} We guess a bounded range of $-10 \le x \le 10$ for the program variable $x$. 
As $i$ remains unchanged within the loop, we guess the assertion map $B$ as 
\(\textstyle
B(\ell) := (-10 \le x \le 10) \wedge (-1 \le i \le 1)
\).
Using an external floating-point analysis tool such as FPTaylor~\cite{fptaylor}, we derive an upper bound on the absolute round-off error of the loop body. Specifically, we provide FPTaylor with the input ranges $x \in [-10,10]$ and $i \in [-1,1]$, together with the loop-body expression $1.5 \cdot x - 0.7 \cdot x + 1.6 \cdot i$. FPTaylor returns a scalar upper bound
\(\textstyle
\gamma_F = 1.847744 \times 10^{-6}
\)
for the absolute round-off error of the update function $F$ for the program variable $x$. For simplicity, We do not consider the bound for the program variable $i$ as it keeps unchanged.
Next, we construct a polynomial template. In this example, we choose degree $d = 2$. Since there are two program variables, $x$ and $i$, the template at location $\ell$ is
\(\textstyle
\eta_\ell = c_0 + c_1 x + c_2 i + c_3 x^2 + c_4 x i + c_5 i^2,
\)
where the coefficients $c_j$ ($0 \le j \le 5$) are unknown and to be resolved.

\smallskip
\noindent{\em Constraint Deriving Stage.} Let $I(\ell)(x,i):=\eta_\ell(x,i)\ge 0$ denote the invariant at the loop entry. We now establish the constraints for $I$ w.r.t the initiation condition~(\ref{eq:initiation}) and the strengthened consecution condition~(\ref{eq:relaxation1}). From the initiation condition, we obtain
\(\textstyle
(-1 \le i \le 1 \wedge x = 0) \implies I(\ell)(x,i).
\)
Applying the first strengthening~(\ref{eq:relaxation1}) to the consecution condition yields
\(\textstyle
(I(\ell)(x,i) \wedge -1 \le i \le 1 \wedge -10 \le x \le 10 \wedge -\gamma_F \le x' - F(x,0) \le \gamma_F)
\implies I(\ell)(x',i).
\) where $F(x,0)=1.5 \cdot x - 0.7 \cdot x + 1.6 \cdot i$.
In this example, the over-approximation $\overline{b}$ is not involved, since the guard condition contains no floating-point arithmetic.

We rewrite the strengthened consecution constraints equivalently as
\begin{align*}\textstyle
& (i + 1 \ge 0 \wedge 1 - i \ge 0 \wedge x + 10 \ge 0 \wedge 10 - x \ge 0 \wedge \label{eq:constraintinex} \\
& \eta_\ell(x,i) \ge 0 \wedge x' - F(x,0) - \gamma_F \ge 0 \wedge F(x,0) + \gamma_F - x' \ge 0)
\implies \eta_\ell(x',i) \ge 0.
\end{align*}
Introducing 
$\mbox{\sl bar} = 0.1$ and $t = 0.0001$, 
we soundly under-approximate this constraint as
\begin{align*}\textstyle
& (i + 1 \ge 0 \wedge 1 - i \ge 0 \wedge x + 10 \ge 0 \wedge 10 - x \ge 0 \\
& \wedge x' - F(x,0) - \gamma_F \ge 0 \wedge F(x,0) + \gamma_F - x' \ge 0)
\implies \eta_\ell(x',i) - \mbox{\sl bar} \cdot \eta_\ell(x,i) \ge t.
\end{align*}
The initiation constraint is handled with $t$ similarly.

\smallskip
\noindent{\em Coefficient Solving Stage.} We now convert these constraints into systems of equations over the unknown coefficients. For the initiation condition~(\ref{eq:inittemplate}), we rewrite it as
\(\textstyle
\forall x,i \; \left[(\forall g \in \Theta,\ g \ge 0) \implies \eta_\ell(x,i) \ge 0\right],
\)
where $\Theta = \{i+1, 1-i, x, -x\}$. Following the sound formulation~(\ref{eq:handelmansum}), we introduce non-negative variables $\lambda_j$ and impose
\(\textstyle
\sum_{j=1}^{s_2} \lambda_j \cdot g_j = \eta_\ell(x,i),
\)
where the polynomials $g_j$ are all products of at most degree-$2$ polynomials from $\Theta$. Expanding both sides yields expressions in the monomials $1, x, i, x^2, x i, i^2$. Matching coefficients of corresponding monomials produces a system of $6$ linear equations involving only the variables $\lambda_j$ and the template coefficients $c_j$. The same application of Handelman’s Theorem is used for the strengthened consecution constraints~(\ref{eq:simplified1}), except that the right-hand side $\eta_\ell(x',i) - \mbox{\sl bar} \cdot \eta_\ell(x,i)$ is a polynomial whose coefficients are linear combinations of the template coefficients. 

To optimize the invariant, we introduce two additional variables, $\mbox{\sl up}$ and $\mbox{\sl low}$, representing the upper and lower bounds of the target variable $x$. By adding the implication $\eta_\ell(x,i) \ge 0 \implies \mbox{\sl low} \le x \le \mbox{\sl up}$, we derive strengthened constraints as in~(\ref{eq:optrelaxed}), and again apply Handelman’s Theorem or Putinar's Positivstellensatz to obtain linear constraints.
Then, we solve the resulting system of equations using an external solver and substitute the computed coefficient values back into $\eta_\ell$, thereby obtaining the candidate invariant. In this example, the candidate invariant is
$\textstyle
0.9999989525954043 - 0.01095421825339196 \cdot x^2 \ge 0,
$
which implies the range $-9.555 \le x \le 9.555$.

\begin{algorithm}
\caption{framework for Bounded floating-point Program}
\label{alg:FloatInvGen-A}

\begin{flushleft}

\textbf{Input:} An fp-CFG $\Gamma$ and an integer polynomial degree $d>0$\\

\textbf{Output:} A valid polynomial invariant $I$ with degree at most $d$ for $\Gamma$

\end{flushleft}

\begin{enumerate}
    \item Guess an assertion map $B$ for bounded ranges of all program variables.

    \item Construct a polynomial invariant template $I := \eta \ge 0$ of degree $d$ over program variables.

    \item Generate the initiation constraint according to~(\ref{eq:initiation}).

    \item For each transition $(\ell, b, F, R', \ell')$ in $\Gamma$:
    \begin{itemize}
        \item Compute an upper bound $\gamma_F$ on the round-off error of the update function $F$ using $\{\mathbf{v} \mid \mathbf{v}\models B(\ell)\}$ as the input range and external round-off error analysis tools.
        \item Generate the consecution constraint~(\ref{eq:consecution}) and strengthen it using $\gamma_F$ to obtain (\ref{eq:relaxation1}) or (\ref{eq:relaxation2}), and then strengthen it with small barrier constant $\mbox{\sl bar}$ and small constant $t$ and get (\ref{eq:simplified1}) or (\ref{eq:simplified2}).
    \end{itemize}

    \item Collect all constraints, add the optimization constraints~(\ref{eq:optrelaxed}), and eliminate program variables to obtain a system of equations over the unknown template coefficients by positivity certificates (e.g.,(\ref{thm:handelman})).

    \item Solve the resulting system using an external solver and substitute the solution back into the invariant template to get a candidate invariant $I$.

    \item If there exists a location $\ell$ such that $I(\ell)$ does not imply $B(\ell)$:
    \begin{itemize}
        \item Enlarge the bounded ranges in $B$.
        \item Rerun Algorithm~\ref{alg:FloatInvGen-A} with the updated $B$ and the same $\Gamma$.
    \end{itemize}

    \item Calculate the aggregated error of the invariant $I$ within $B$ w.r.t initiation and consecution constraints. If the error is larger than the additional strengthening constant $t$, then return \textbf{False}.
    \item Return the candidate invariant $I$.
\end{enumerate}
\end{algorithm}

\smallskip
\noindent{\em Validation Stage.} Finally, we check whether the candidate invariant implies the guessed range for the program variable $x$, which in this case is $-10 \le x \le 10$. This clearly holds. We do not need to check the variable $i$, as it remains constant and is not involved in the invariant, and always stays within its initial range in $B$. Moreover, we verify that the accumulated numerical error over the range $-10 \le x \le 10$ introduced by the external solver, which is $3.79 \times 10^{-7}$, is smaller than $t=0.0001$. Therefore, we conclude that the candidate invariant is valid.

\subsection{Soundness Arguments}\label{sect:soundness}

The main concern of soundness of our framework comes from the potentially invalid guessed ranges in $B$ which affect the strengthening with round-off bounds and over-approximation of guard conditions. This is addressed in our validation stage.  
The following is a detailed proof.

\begin{theorem}
If Algorithm~\ref{alg:FloatInvGen-A} returns an assertion map $I$, then $I$ is a valid invariant.
\label{thm:soundness}
\end{theorem}

\begin{proof} 
The main point is to show that if $B$ passes the validation stage, then $B$ is indeed an invariant. We prove this as an intermediate part in an inductive proof.
Let $(\ell_0,\mathbf{v}_0),\dots,(\ell_n,\mathbf{v}_n)$ be any path of the input fp-CFG $\Gamma$. We prove by induction on $k$ that each program state $(\ell_k,\mathbf{v}_k)$ ($0\le k\le n$) fulfills that $\mathbf{v}_k\models I(\ell_k)$.
Recall that from the validation stage, 
we have \( I(\ell) \models B(\ell) \) for all locations $\ell$. 
We first present the proof assuming no errors from external solvers, and then account for these errors via numerical repair.

\smallskip
\noindent{\textbf{Base Case:}} At the start of the input fp-CFG, we have that the initial condition $\mbox{\sl Init}$ implies $B(\ell^*)$ as well as $I(\ell^*)$.
Therefore, we have $\mathbf{v}_0\models I(\ell^*)$. 

\smallskip
\noindent{\textbf{Inductive Step:}} Suppose that $\mathbf{v}_k\models I(\ell_k)$ for $k<n$. 
Since $I(\ell_k)$ implies $B(\ell_k)$, we have $\mathbf{v}_k\models B(\ell_k)$. Let the transition taken from $(\ell_k, \mathbf{v}_k)$ to $(\ell_{k+1}, \mathbf{v}_{k+1})$ be $(\ell_k, b, F, R',\ell_{k+1})$.
We prove that $\mathbf{v}_{k+1}\models I(\ell_{k+1})$. Consider the consecution condition (\ref{eq:consecution}). First, since $\mathbf{v}_k\models B(\ell_k)\wedge \langle b\rangle$, we have  $\mathbf{v}_k\models \overline{b}$ as $\langle b\rangle$ implies $\overline{b}$ given $B(\ell_k)$ (addressing the soundness of over-approximation of guard conditions). Moreover, we have  $|\mathbf{v}_{k+1} - F(\mathbf{v}_{k},\mathbf{0})|\le \gamma_F$ (i.e., the round-off bounds in $\gamma_F$ are valid). Second,
as (i) the invariant $I$ fulfills the strengthened condition of either (\ref{eq:relaxation1}) or (\ref{eq:relaxation2}) and (ii) $\mathbf{v}_{k+1}=F(\mathbf{\mathbf{v}_k},\mathbf{r})$ for some error valuation $\mathbf{r}$ overall deviation bounded by $\gamma_F$, we have $\mathbf{v}_{k+1}\models I(\ell_{k+1})$ 
(addressing the soundness of strengthened consecution conditions). Therefore, both $I$ and $B$ are invariants. 

When the numerical errors of external solvers are considered, at the base step and each inductive step, these numerical errors are covered by the strengthening constant
$t>0$ within the bounded ranges of $B$ (see Appendix~\ref{sec:NR}). 
\qed
\end{proof}

\begin{remark}
Iterative guessing of the initial ranges is the main weakness of our framework. However, this is mitigated by the following practical considerations. First, floating-point programs often have bounded program variables, so that valid ranges can eventually be guessed by iterative enlarging. Second, although enlarged ranges cause larger round-off bounds, the magnitude of such round-off bounds is much smaller than the numerical values appearing in the program. This suggests that the accuracy (tightness of generated invariants) of our framework is insensitive to larger round-off bounds caused by enlarged guessed ranges. \qed
\end{remark}

\begin{remark}
Our framework handles unstable conditionals 
caused by round-off errors,
as all floating-point guard conditions $\langle b \rangle$ are relaxed to their over-approximations $\overline{b}$ such that $\langle b \rangle \models \overline{b}$. As a result, all execution paths that are valid in the floating-point model are preserved. The converse implication, $\overline{b} \models \langle b \rangle$, may not hold in general. Consequently, some infeasible paths that become feasible only due to pessimistic round-off errors when the original guard $\langle b \rangle$ is false -- may satisfy $\overline{b}$ however -- are included in the analysis. Since the computed invariant is an over-approximation, including such additional paths does not affect soundness.\qed
\end{remark}

\section{Evaluation}\label{sec:exp}

We evaluate our framework by implementing the Algorithm~\ref{alg:FloatInvGen-A} with strategy~(\ref{eq:relaxation1}) with direct optimization way (denoted \textbf{S1}) and strategy~(\ref{eq:relaxation2}) with divide-and-conquer optimization way (denoted \textbf{S2}) and compare them with the state-of-the-art tools over a collection of loop benchmarks involving polynomial computations and divisions. Our evaluation addresses two questions:
\begin{itemize}
  \item[\textbf{RQ1}] How does our prototype compare to state-of-the-art tools in terms of invariant precision and runtime?
  \item[\textbf{RQ2}] How does the choice of polynomial degree affect effectiveness and cost?
\end{itemize}
To answer \textbf{RQ1}, we compare against FPTaylor~\cite{fptaylor}, PINE~\cite{pine}, and TVPI-FP~\cite{tvpifp}. FPTaylor uses a first-order differential characterization to derive variable ranges, PINE is data-driven and samples program states to fit ellipsoidal invariants, and TVPI-FP uses a two-variable affine domain for floating-point invariants. We compare the precision of generated invariants by examining ranges of key program variables (as in~\cite{tvpifp}).
To address \textbf{RQ2}, we vary the polynomial template degree in Algorithm \ref{alg:FloatInvGen-A} and measure accuracy and runtime. 

\subsection{Experiment Setup}

\smallskip
\noindent{\em Implementation.} We implement both strengthening (\textbf{S1}, \textbf{S2}) with different optimization ways in Python 3.12 and use PySMT~\cite{pysmt} for expression handling, AMPLpy~\cite{amplpy} to call the optimizer Gurobi~12.0.2~\cite{gurobi}, and FPTaylor~\cite{fptaylor} to compute upper bounds \(\delta_F\) for absolute round-off errors. Gurobi is run with minimum tolerance (maximum precision). We use $\mbox{\sl bar}=0.1$ and $t=0.0001$ in (\ref{eq:simplified1}) and (\ref{eq:simplified2}) and apply Handelman's theorem to convert positivity constraints into linear systems for the solver. We use Z3 \cite{z3} in the final validation stage for the guessed ranges in $B$.

\smallskip
\noindent{\em Benchmarks.}
We select loop benchmarks from FPBench~\cite{fpbench}, TVPI-FP~\cite{tvpifp}, PINE~\cite{pine}, FLUCTUAT\cite{review2}, and verif-iqc
\cite{review1}: all loop benchmarks from TVPI-FP (excluding a non-reachable loop `springf'), representative PINE examples, and eight FPBench benchmarks (excluding those e with other suites). No benchmark triggers overflow/underflow; none has an unavoidable division-by-zero. When FPTaylor raises a division-by-zero possibility (e.g., Raphson), we add a pre-check that halts when the divisor is zero so fo-DC applies. For infinite-loop benchmarks, we craft finite-loop variants with a fixed iteration number to allow comparison with 
FPTaylor (which struggles with unbounded loops). All experiments use single-precision floating-point (PINE's default); note that tools may differ slightly in \(\epsilon\) settings (e.g., FPTaylor uses \(\epsilon=2^{-24}\), TVPI-FP uses \(\epsilon=2^{-23}\)) which are not configurable. 

\smallskip
\noindent{\em Machine.} Most experiments run on an M4 MacBook Air (16\,GB RAM) single-threaded, except for that TVPI-FP  
is executed separately on an Ubuntu VM (i7-13650HX, 2.60\,GHz) (since it runs only on x86); hence we do not compare TVPI-FP runtimes.

\smallskip
\noindent{\em Initial Guessed Range.} The initial range estimate is provided as input to our algorithm, and the time required to generate this estimate is excluded from our reported results. We obtain this estimate via a data-driven approach and subsequently soundly enlarge it to ensure coverage of the observed values. Any result that passes the verification step guarantees the validity of the inferred range.

\subsection{Experiment Results}

\smallskip
\noindent\textbf{Answering RQ1.} For most comparison we set template degree and Handelman multiplicand bound \(m\) both to 2 for most benchmarks; for SineNewton and Raphson, we set degree and $m$ to 4 as 2-degree template fails to get meaningful results. TVPI-FP runtime is omitted due to hardware differences.

According to Table~\ref{tab:experiment_results}, our approach solves all but two benchmarks and produces the tightest range on 22 of 27 benchmarks. Typical runtimes are modest (seconds). PINE (with its updated SMT backend) is fast on some instances but fails on many benchmarks; compared to PINE's original reported results, our invariants are generally tighter. FPTaylor handles bounded-iteration loops well but fails on unbounded loops; on bounded loops our results are often comparable or tighter. TVPI-FP solves fewer benchmarks but handles some complex cases (e.g., RateLimiter, \texttt{sqrt1}). Both of our methods and the baselines fail on \texttt{nBodySimulation} (infinite range). \textbf{S1} typically yields good bounds quickly, while \textbf{S2} can give tighter bounds at the cost of longer solve time. The experimental results also show that a small choice of degree suffices to handle benchmarks with complex polynomial and division operations. 

Note that FLUCANT is a commercial tool, and therefore we cannot perform a direct comparison. However, based on the data reported in~\cite{review2}, our result for \texttt{order2FilterLinear} is better than the bound obtained by FLUCANT with 60 unfolding iterations, and only slightly worse than that obtained with 100 iterations. For the \texttt{order2FilterUncertainty} case, our result outperforms FLUCANT even with 100 unfolding iterations. We also do not directly compare our results with verif-iqc~\cite{review1}, as their approach requires a tight invariant as a prerequisite. Nevertheless, according to the data reported in their paper, our results are only slightly more conservative than the bounds they obtain.
\begin{table}[ht]
    \centering
    \resizebox{\textwidth}{!}{
    \begin{tabular}{@{}lcc|cc|cc|cc|c@{}}
        \toprule
        \textbf{Benchmark} 
        & \multicolumn{2}{c|}{\textbf{S1}} 
        & \multicolumn{2}{c|}{\textbf{S2}} 
        & \multicolumn{2}{c|}{\textbf{PINE}} 
        & \multicolumn{2}{c|}{\textbf{FPTaylor}} 
        & \multicolumn{1}{c}{\textbf{TVPI-FP}} \\ 
        \cmidrule(lr){2-3} \cmidrule(lr){4-5} \cmidrule(lr){6-7} \cmidrule(lr){8-9} \cmidrule(lr){10-10}
        & \textbf{Range} & \textbf{Time (s)} 
        & \textbf{Range} & \textbf{Time (s)} 
        & \textbf{Range} & \textbf{Time (s)} 
        & \textbf{Range} & \textbf{Time (s)} 
        & \textbf{Range} \\ \midrule
        SineNewton* & \textbf{1.84} & 1.6  & F      & -    & \textbf{2.0} & 2.81     & F      & -      & F \\ 
        \textbf{SineNewton(10 iter)}* & \textbf{1.57} & 12.29 & F & - & 2.0 & 2.81 & F & - & F \\ 
        PendulumSmall(100 iter) & 2.97 & 0.72 & 2.97 & 9.87 & F & - & \textbf{2.87} & $<0.0001$ & F \\ 
        \textbf{PendulumSmall} & 2.97 & 0.52 & \textbf{2.88} & 8.71 & F & - & F & - & F \\ 
        ex1 & 19.11 & 0.34 & \textbf{16.0} & 10.94 & F & - & F & - & F \\ 
        ex2 & 4.8 & 0.44 & \textbf{2.0} & 10.4 & 2.2 & 3.01 & F & - & F\\ 
        leadlag(1000 iter) & 7.28 & 8.2 & \textbf{5.04} & 11.26 & F & - & 84668.667 & $<0.0001$ & F\\ 
        \textbf{leadlag} & \textbf{6.78} & 3.53 & 6.79 & 12.21 & F & - & F & - & F\\ 
        gaussian(100 iter) & 8.69 & 4.04 & \textbf{8.21} & 25.93 & F & - & 9.152 & $<0.0001$ & F\\ 
        \textbf{gaussian} & 8.69 & 12.82 & \textbf{8.64} & 15.98 & F & - & F & - & F\\ 
        coupledMass(10 iter) & 48.17 & 38.33 & \textbf{20.60} & 36.09 & F & - & 86.45 & $<0.0001$ & F\\ 
        \textbf{coupledMass} & TO & - & \textbf{30.54} & 24.2 & F & - & F & - & F\\ 
        dampened(200 iter) & 10.45 & 3.14 & \textbf{2.64} & 9.49 & F & - & 163.28 & $<0.0001$ & F \\ 
        dampened & TO & - & \textbf{2.66} & 10.93 & F & - & F & - & F \\ 
        harmonic(200 iter) & 32.3 & 2.92 & \textbf{25.74} & 48.99 & F & - & 25.8 & $<0.0001$ & F\\ 
        \textbf{harmonic} & 55.24 & 1.59 & \textbf{25.73} & 42.94 & F & - & F & - & F\\ 
        butterworth(10 iter) & TO & - & 2.90 & 10.84 & F & - & \textbf{2.83} & $<0.0001$ & F\\
        butterworth & TO & - & 2.90 & 24.05 & F & - & \textbf{2.83} & $<0.0001$ & F\\ 
        RateLimiter & \textbf{259.74} & 2.54 & F & - & F & - & F & - & 278.50 \\ 
        sqrt1* & TO & - & F & - & 1.1 & 0.26 & F & - & \textbf{0.37}\\
        artificial & \textbf{5.0} & 0.34 & \textbf{5.0} & 8.05 & \textbf{5.0} & 0.15 & F & - & \textbf{5.0}\\
        bigLoop & \textbf{$<0.001$} & 0.81 & \textbf{$<0.001$} & 9.02 & F & - & \textbf{$<0.001$} & $<0.0001$ & 10\\
        \textbf{nBodySimulation}* & F & - & F & - & F & - & F & - & F\\
        \textbf{Raphson}* & \textbf{$<0.0001$} & 1.66 & F & - & F & - & F & - & F \\
        LTI & \textbf{5.6} & 5.49 & F & - &F & -& F & - &F \\
         order2FilterLinear & \textbf{5.306} & 0.61 & 6.2 & 1.33 &F & -& F & - &F \\
          order2FilterUncertainty & \textbf{27.556} & 2.43 & F & - &F & -& F & - &F \\
        \bottomrule
    \end{tabular}
    }
    \caption{Experiment results. The columns list benchmark names (FPBench benchmarks in bold; benchmarks with division marked with `*'), S1 and S2 results (range length \(\mbox{\sl up}-\mbox{\sl low}\) and solve time), and the results from PINE, FPTaylor, and TVPI-FP. `TO' denotes timeout; `F' denotes that the solver finished but failed to find an invariant; bold indicates the tightest range.}
    \label{tab:experiment_results}
\end{table}

\smallskip
\noindent{\bf Answering \textbf{RQ2}.} We explore higher-degree templates (degrees $4$ and $7$) to measure accuracy gains with a longer time budget. Using \textbf{S1} only (\textbf{S2} would require multiple costly solves for binary search), we run a 30-minute time limit and report results in Table~\ref{tab:da}. Degree increases generally raise term size and solver time; many benchmarks time out at higher degree. When solutions are found, higher degree improves tightness (notably SineNewton and ex2). Thus higher degree can yield more precise invariants, but at heavy computational cost.

\begin{table}[ht]
    \centering
    \resizebox{\textwidth}{!}{
    \begin{tabular}{@{}lccc|ccc|ccc@{}}
        \toprule
        \textbf{Benchmark} & \textbf{degree=2} & & &\textbf{degree=4} & & &\textbf{degree=7} \\ 
        \cmidrule{2-2} \cmidrule{5-5} \cmidrule{8-8} 
        & \textbf{Range} & \textbf{Time (s)} & \textbf{Term Size} & \textbf{Range} & \textbf{Time (s)} & \textbf{Term Size} & \textbf{Range} & \textbf{Time (s)} & \textbf{Term Size}\\ \midrule
        SineNewton(10 iter) & 2.0 & 1.05 & 42 & 1.57 & 12.29 & 265 & 0.13 & 337.02 & 1920\\ 
        PendulumSmall(100 iter) & 2.97 & 0.63 &  88 &TO  &  - & 4794 & TO &  - &  20071 \\
        PendulumSmall & 2.97 & 0.46 & 57 & TO &- & 2942& TO & - & 6638\\
        ex1 & 17.99 & 13.50 & 40 & 16.21 & 92.02& 691 & TO& -& 1919 \\
        ex2 & 4.8 & 0.33 & 41 & 2.5 & 44.69 & 511 & TO &  - & 2044 \\
         leadlag(1000 iter) & 7.28 & 7.33 & 125 & TO & -& 7428 & - & - & 51935\\ 
         leadlag & 6.78 & 4.78 & 88 & TO &- &4809 & TO & - & 20071\\ 
         gaussian(100 iter) & 8.69 & 5.56 & 201 & TO & -& 27819& TO& -& 248552\\ 
         gaussian & 8.68& 12.74 & 154 & TO& -& 20402& TO & -& 246704\\ 
        coupledMass(10 iter) & 48.17 & 38.33 & 369 & TO & -& 246516 & TO&- &$>1000000$\\  
        dampened(200 iter) & 10.45 & 3.14 & 91 & TO & - &4885 & TO & - & $>1000000$\\
         harmonic(200 iter) & 32.3 & 2.92 & 91 & TO& - &4706 & TO & - &$>1000000$ \\ 
          harmonic& 55.24 & 1.59 & 58 & TO &- & 2918 & TO& -& 19777\\ 
           RateLimiter & 259.74 & 2.54 &118 & 256.23 & 478.37 &2354 & TO& -&13334 \\
           Raphson & F & - & 25 & <0.0001 & 1.66 & 105 & <0.0001 & 56.27 & 489\\
            LTI & 5.6 & 5.49 & 149 & TO & - &20933 & TO & - &  249351\\
         order2FilterLinear & 5.306 & 0.61 &102 & TO & - & 3965 &TO & - & $>1000000$ \\
          order2FilterUncertainty & 27.556 & 2.43 & 69 & TO & - & 2229 & TO& - & 563823\\
        \bottomrule
    \end{tabular}
        
    }
    \caption{\textbf{S1} with different degrees. 'Term Size' denotes the number of fresh variables \(\lambda_j\) in the Handelman-based constraint system \(\mathcal{C}_{\mathrm{handelman}}\).`TO' denotes timeout; `F' denotes failure to find an invariant.}
    \label{tab:da}

\end{table}

\begin{remark}
Abstract-interpretation tools such as Astrée~\cite{BCC+03} and Fluctuat~\cite{Gou01,GMP02} also target floating-point programs. They are not publicly available for comparison, so we provide a qualitative contrast. Fluctuat focuses on round-off propagation (e.g., via zonotopes) and can bound both floating-point ranges and the deviation from exact real-valued semantics. Its emphasis is error tracking rather than invariant generation. Our method targets invariant generation without introducing explicit error symbols; we do not attempt to bound the real–float discrepancy. Astrée linearizes floating-point expressions into real-number forms with interval coefficients and then applies classical linear domains. This approach is general and efficient, but infers only linear invariants and loses precision when converting interval-coefficient expressions into scalar-coefficient forms. Our method trades some generality for more precise polynomial invariants in many benchmarks. \qed
\end{remark}

\section{Discussion to Avoid Guessing Bounded Ranges}

We discuss how our framework can be modified to circumvent the iterative guess of bounded range but is limited to polynomial floating-point programs without division. 
Due to the space constraints, the details is relegated to Appendix~\ref{sec:al2ap}. 


The key idea is to replace the use of constant round-off bounds with symbolic bounds for round-off errors. Instead of guessing bounded ranges $B$ during the preparation stage, we introduce, for each program variable $x$, an absolute-value variable $x_{abs}$ together with the constraints: $\textstyle x^2 = x_{ abs}^2$ and $\textstyle x_{abs} \ge 0$. 
These variables symbolically represent the absolute values of program variables and allow us to express error bounds without assuming concrete ranges.

For each transition $(\ell, b, F, R', \ell')$, 
we apply the fo-DC method~(\ref{eq:fodc}) to the update function $F$ and obtain the Taylor expansion:
$\textstyle
\textstyle\left| F(\mathbf{x}, \mathbf{r}) - F(\mathbf{x},\mathbf{0}) \right|
\le
\sum_i \left| \frac{\partial \hat{F}}{\partial r_i} (\mathbf{x}, \mathbf{0}) \right| \cdot \epsilon_i
+ \left| R_2(\mathbf{x}, \mathbf{r}) \right|
$, 
where each $\epsilon_i$ equals $\epsilon$ if $r_i$ is a relative error variable and $\delta$ if it is an absolute error variable. Since we assume no division in assignments, the partial derivatives $\textstyle\frac{\partial \hat{F}}{\partial r_i} (\mathbf{x}, \mathbf{0})$ are polynomials. This allows us to bound their absolute values by repeated application of the following equality and triangle inequality:
$\textstyle\left| c \cdot \prod_i x_i \right| = |c| \cdot \prod_i x_{\mbox{\sl abs}}
\quad \text{and} \quad
|f + g| \le |f| + |g|$.
Similarly, we bound the second-order term $\textstyle\left| R_2(\mathbf{x}, \mathbf{r}) \right|$.
The resulting upper-bound vector $\gamma_F$ is therefore a vector of polynomials over the absolute-value variables $x_{\mbox{\sl abs}}$, providing symbolic bounds for 
round-off errors without guessing bounded ranges.

\section{Related Work}

We compare our approach with most-related floating-point analysis results in the literature. Further works on round-off error analysis are discussed in Appendix~\ref{sec:Round-off}. 


\emph{Abstract interpretation} \cite{CC77-1} is a classical method that provides a unified framework for invariant generation. The abstract interpretation based static analyzer \emph{ASTR\'EE} \cite{BCC+03} employs the floating-point abstraction technique of \cite{Min04} to soundly abstract floating-point expressions into ones over real numbers, and then infers linear invariants using conventional abstract domains that are designed for real-number semantics. FLUCTUAT \cite{Gou01,GMP02} bounds the errors due to the finite precision implementation and traces the source of errors in floating-point programs based on abstract interpretation. Recently, \emph{TVPI-FP} \cite{tvpifp} establishes a two-variable affine abstract domain for analyzing floating-point programs, by using interval coefficients in the inequalities. 

\emph{Pine} \cite{pine} is a data-driven approach that first samples program executions and then seeks to enclose the sampled points by ellipsoid invariants (a subclass of polynomial invariants) whose correctness is validated by SMT solvers. The approach is lightweight in the sense that the complexity of the program itself does not significantly influence its performance, but the data-driven nature also introduces a considerable amount of randomness. In our approach, the guessing of an initial assertion map could follow such approach.

\emph{FPTaylor} \cite{fptaylor} is primarily used to calculate floating-point round-off bounds rather than invariants. It utilizes a simple interval to determine the range of each program variable, allowing it to find results for finite-loop cases efficiently. However, this interval approach tends to ignore the relationships between variables, resulting in invariants that are generally too coarse for broader applications.

\emph{Code-level formal verification of invariant sets.} The work in~\cite{review2} verifies ellipsoidal invariants of linear parameter-varying (LPV) systems at the code level by combining control-theoretic invariant generation with deductive verification in Frama-C. It translates invariant certificates into ACSL annotations and proves them using SMT solvers, while accounting for floating-point errors via sound over-approximations. This approach provides end-to-end guarantees from model to implementation, but is tailored to systems with known ellipsoidal invariants.


Compared with the approaches above, our approach has its root in polynomial constraint solving~\cite{DBLP:conf/pldi/AsadiC0GM21,polyqent,DBLP:journals/toplas/WuWXZZY25,DBLP:conf/cav/ColonSS03,DBLP:conf/sas/SankaranarayananSM04} for invariant generation, and therefore is orthogonal. It is also worth noting that existing polynomial solving methods consider exact real arithmetic and ignore round-off errors, while our approach carefully handles round-off errors. 

\section{Conclusion and Future Work}

We propose a novel framework for generating polynomial invariants in floating-point programs by polynomial constraint solving. Our key contribution is integrating polynomial solving with round-off error analysis 
to eliminate error variables via round-off bounds,   
enhanced by barrier certificates, numerical repair constant, and optimization techniques. Experiments on diverse benchmarks show that our approach produces tighter invariants than state-of-the-art methods, especially for programs with complex polynomial and division computations. An important future work is to extend our framework to handle transcendental functions, which would require to investigate detailed implementation behind and proper abstractions for transcendental functions. 


\section*{Data Availability Statement }
The results of this study are available at \url{https://doi.org/10.5281/zenodo.19812670}.

\bibliographystyle{unsrt}  
\bibliography{reference}  

\newpage
\appendix
\section{Round-off error analysis tools}\label{sec:Round-off}
Round-off error analysis is an orthogonal topic to invariant generation as mentioned previously.


FPTaylor \cite{fptaylor} uses symbolic Taylor expansion to estimate the maximum floating-point round-off error in straight-line code. It approximates the floating-point expression with its first order Taylor expansion, and simply uses a coarse bound for the second order term.  
Then it employs rigorous global optimization to compute error bounds. 

PRECiSA \cite{MoscatoTDM17,TitoloFMM18,TitoloMFMM24} represents errors of floating-point operations symbolically in
terms of the errors of the operands relative to the real-valued operations, and then solves the collected constraints via branch-and-bound techniques. It also propagates conditional error bounds into function calls and branches.

Satire \cite{DasBGKP20} also implements the symbolic Taylor expression-based approach, with additional optimizations for efficiency. However, some of these optimizations, such as dropping higher-order terms, may make the analysis unsound.

Daisy \cite{DarulovaINRBB18} provides a framework for accuracy analysis and synthesis of numerical programs, which integrates several different sound static analyses and optimizations for floating-point and fixed-point programs, covering absolute and relative rounding errors for arithmetic. Recently, Daisy also handles function calls~\cite{AbbasiD23} and array-like data structures~\cite{IsychevD23}.

Real2Float \cite{MagronCD17} over-approximates absolute round-off errors of floating-point nonlinear programs,  using optimization techniques including semidefinite programming and sums of squares certificates.

Gappa \cite{DM10} uses interval arithmetic and forward error analysis to certify the bounds for ranges and rounding errors of floating-point expressions. However, it currently applies to only straight-line floating-point programs.

VCFloat \cite{RMML16} and VCFloat2\cite{AK24} provide semi-automated floating-point round-off error analysis, by generating an annotated real-number expression with appropriate error bounds from a floating-point expression, and then using Coq to reason over the real-number expression.

Rosa~\cite{DarulovaK14} encodes reasoning over roundoff errors into that over real numbers, and then combines SMT solving with affine and interval arithmetic for computing over-approximation of round-off errors. 

\section{A Framework to Avoid Iterative Guess of Bounded Ranges}\label{sec:al2ap}
In the framework, we still apply the fo-DC, but we no longer derive a constant upper bound vector $\gamma_F$ for the absolute rounding error due to the absence of a bounded initial assertion map. The key point to address this difficulty is to directly have symbolic over-approximation over the fo-DC. 

The detailed workflow of the algorithm for the input fp-CFG  $\Gamma=(L, X, R, \mbox{\sl Init}, \rightarrow)$ is as follows.

\smallskip
\noindent{\em Step 1: Fresh variables for absolute program values.} For each program variable $x\in X$ in $\Gamma$, we have a fresh variable $x_{\mbox{\sl abs}}$ that represents the absolute value of $x$. We have two polynomial constraints 
\begin{equation}\label{eq:absconstraints}
x^2 = x^2_{\mbox{\sl abs}}\mbox{ and }x_{\mbox{\sl abs}}\ge 0 
\end{equation}
to ensure that indeed $x_{\mbox{\sl abs}}$ corresponds to the absolute value of $x$. These fresh variables are the key to derive our symbolic over-approximation for the absolute round-off error.




\smallskip
\noindent{\em Step 2: Polynomial symbolic upper bounds for absolute rounding errors for normalized float.} For each transition $(\ell, b, F, R', \ell')$ of $\Gamma$, our algorithm applies the fo-DC in (\ref{eq:fodc}) to the update function $F$ and obtains the Taylor expansion below
\vspace{-0.1cm}
\[\left| F(\mathbf{x}, \mathbf{r}) - F(\mathbf{x},\mathbf{0}) \right| \le   \sum_{i} \left| \frac{\partial \hat{F}}{\partial r_i} (\mathbf{x}, \mathbf{0}) \right|\cdot \epsilon_{i}  + \left|R_2(\mathbf{x}, \mathbf{r}) \right|
\]
where each $\epsilon_i$ is either $\epsilon$ if $r_i$ (the $i$th coordinate of $\mathbf{r}$) is a relative error variable or $\delta$ if $r_i$ is an absolute error variable. As the input program does not have division, the expressions $\frac{\partial \hat{F}}{\partial r_i} (\mathbf{x}, \mathbf{0})$ are polynomial, so that we bound the absolute values of these polynomials  by triangle inequalities:
\[
\left|c\cdot \prod_i x_i\right| = |c|\cdot \prod_i x_{\mbox{\sl abs}}\mbox{ and }|f + g|\le |f| + |g|. 
\]
\vspace{-0.1cm}
We also bound the second-order term $\left|R_2(\mathbf{x}, \mathbf{r}) \right|$ by the triangle inequality, while replacing the absolute value of each $|r_i|$ by its corresponding maximum error (either $\epsilon$ or $\delta$). The resultant upper bound vector $\gamma_F$ for $\sum_{i} \left| \frac{\partial \hat{F}}{\partial r_i} (\mathbf{x}, \mathbf{0}) \right|\cdot \epsilon_{i}  + \left|R_2(\mathbf{x}, \mathbf{r}) \right|$ is a vector of polynomials in the absolute versions $x_{\mbox{\sl abs}}$ of program variables.  



\smallskip
\noindent{\em Step 3: Strengthened constraints for the consecution condition.} By using the symbolic upper bounds $\gamma_F$ derived from the previous step, we first relax $\langle b\rangle$ into $\overline{b}$ similar to what we do in Section~\ref{sec:DA} Relaxation of guard conditions step.

To be specific, for $b = \alpha_1\le \alpha_2$, we derive the symbolic polynomial upper bound $\gamma_{F_i}$ ($i=1,2$) for the absolute rounding error of $F_i$ as in Step 2 of the algorithm, and obtain the over-approximation $\overline{b}:=\alpha_1 - \gamma_{F_1} \le \alpha_2 + \gamma_{F_2}$ again. The over-approximation $\overline{b}$ in other cases is handled in the same way as stated previously. 

Then, we have two strengthening for the consecution condition analogous to our first algorithm. Below we consider a transition $(\ell, \overline{b}, F, R', \ell')$ of $\Gamma$.

\smallskip
\noindent\emph{First strengthening.} The first strengthening for the transition introduces fresh variables $\mathbf{x}'$ to represent the program values after the transition. With $\gamma_F$, we strengthen the consecution condition in almost the same form as (\ref{eq:relaxation1}) as in our first algorithm:
\vspace{-0.1cm}
\begin{equation}\label{eq:relaxation3}
\forall \mathbf{x},\mathbf{x}',\mathbf{x}_{\mbox{\sl abs}}. 
\left[\left(\mathbf{x}\models (I(\ell) \wedge  \overline{b}) \wedge -\gamma_F \le \mathbf{x}' - F(\mathbf{x},\mathbf{0})  \le \gamma_F\wedge \mbox{\sl abs}(\mathbf{x},\mathbf{x}_{\mbox{\sl abs}}) \right) \Rightarrow \mathbf{x}'\models I(\ell')\right]
\end{equation}
where the predicate $\mbox{\sl abs}(\mathbf{x},\mathbf{x}_{\mbox{\sl abs}})$ is 
$\bigwedge_{x\in X} (x^2 = x^2_{\mbox{\sl abs}}\wedge x_{\mbox{\sl abs}}\ge 0)$.
The only difference in the above strengthening is that $\gamma_F$ is now a vector of polynomials in variables $x_{\mbox{\sl abs}}$, and we have $\mbox{\sl abs}(\mathbf{x},\mathbf{x}_{\mbox{\sl abs}})$ to enforce constraints on absolute-value variables.

\smallskip
\noindent\emph{Second strengthening.} The second strengthening for the  transition uses the same $\gamma_F$ as in the first above, introduces the same fresh variables $\mathbf{r}_X$ as in our first algorithm, and strengthens the consecution condition in the same form of (\ref{eq:relaxation2}):
\vspace{-0.1cm}
\begin{equation}\label{eq:relaxation4}
\forall\mathbf{x},\mathbf{x}_{\mbox{\sl abs}},\mathbf{r}_X. \left[\left((\mathbf{x}\models I(\ell)\wedge  \overline{b}) \wedge -\gamma_F \le \mathbf{r}_X\le \gamma_F\wedge \mbox{\sl abs}(\mathbf{x},\mathbf{x}_{\mbox{\sl abs}})\right)\Rightarrow F(\mathbf{x},\mathbf{0}) + \mathbf{r}_X\models I(\ell')\right].
\end{equation}

The correctness of the above strengthenings is illustrated as follows. Note that we no longer require the correctness of  a guessed assertion map of bounded ranges. 

\begin{proposition}\label{prop:relaxation34}
Both the strengthened constraints in (\ref{eq:relaxation3}) and (\ref{eq:relaxation4}) implies the original consecution constraint (\ref{eq:consecution}).
\end{proposition}
\begin{proof}
The proof follows directly from the correctness of the upper bounds $\gamma_F$ and the over-approximations $\overline{b}$.
\end{proof}

A pseudo-code of the algorithm is given below, followed by the soundness argument. 

\begin{algorithm}[H]

\begin{flushleft}
\textbf{Input:} A floating-point control flow graph (fp-CFG) $\Gamma$ with only polynomial expressions and a integer d \\
\textbf{Output:} An invariant $I$
\end{flushleft}

\begin{enumerate}
    \item For each transition $(\ell, b, F, R', \ell')$ of $\Gamma$:
    \begin{itemize}
        \item Introduce absolute variables for each fresh variable in $F$ and store them in $X_{\mbox{abs}}$.
        \item Introduce absolute variable constraints for $X_{\mbox{abs}}$ following (\ref{eq:absconstraints})

        \item Compute the polynomial symbolic upper bounds $\gamma_F$ as described in Step 2 of Section~\ref{sec:al2ap} using $X_{\mbox{abs}}$.
        
        \item Compute the over-approximation $\overline{b}$ by the $\gamma_F$'s for the arithmetic expressions in $b$.
    
        \item With $\gamma_F$ and $\overline{b}$, strengthen the consecution condition for the transition with (\ref{eq:relaxation3}) or (\ref{eq:relaxation4}).

    \end{itemize}

    \item Solve the initiation, the strengthened consecution, and the absolute variable constraints as in Section ~\ref{sec:DA} Coefficient Solving Stage to obtain the final invariant $I$.

    \item Return $I$.
\end{enumerate}
\end{algorithm}


\begin{theorem}
If this algorithm returns an assertion map $I$, then $I$ is a correct invariant for the input fp-CFG. 
\end{theorem}
\begin{proof}
The result follows directly from Proposition~\ref{prop:relaxation34} and the correctness of invariant solving in Section~\ref{sect:soundness}.
\end{proof}


\end{document}